\documentclass[12pt,preprint]{aastex}

\newcommand{\Msun}{\mbox{ M}_{\odot}}
\newcommand{\gae}{\mathrel{>\kern-1.0em\lower0.9ex\hbox{$\sim$}}}
\newcommand{\lae}{\mathrel{<\kern-1.0em\lower0.9ex\hbox{$\sim$}}}

\begin{document}

\title{Chandra Grating Spectroscopy of the X-ray Binary 4U 1700-37 in a 
Flaring State}

\author{Bram Boroson}
\affil{Center for Space Research, Massachusetts Institute of Technology, 
Cambridge, MA 02138; bboroson@space.mit.edu}

\and

\author{Saeqa Dil Vrtilek}
\affil{Harvard-Smithsonian Center for Astrophysics, 60 Garden Street,
Cambridge, MA 02138; svrtilek@cfa.harvard.edu}

\and

\author{Timothy Kallman and Michael Corcoran}
\affil{NASA Goddard Space Flight Center; Greenbelt, MD 20771
tim@xstar.gsfc.nasa.gov, corcoran@barnegat.gsfc.nasa.gov}

\begin{abstract}

Chandra X-ray Observatory grating spectra of the supergiant X-ray Binary
4U~1700-37 reveal emission lines from hydrogen and helium-like S, Si, Mg,
and Ne in the 4--13 \AA\ range.  The spectrum also shows fluorescent lines
from S, Si, and a prominent Fe~K$\alpha$ line at 1.94\AA.  The lines
contribute to the previously unaccounted ``soft excess'' in the flux in
this range at orbital $\phi\approx0.7$.  The X-ray source was observed
during intermittent flaring, and the strengths of the lines vary with the
source state. The widths of the lines (FWHM $\approx1000-2000$~km/s) can
result from either Compton scattering or Doppler shifts. Power spectra of
the hard X-rays show red noise and the soft X-rays and lines show in
addition quasiperiodic oscillations (QPOs) and a power-spectral break.
Helium-like triplets of Si and Mg suggest that the gas is not in a pure
photoionization equilibrium.  We discuss whether resonant scattering could
affect the line ratios or whether a portion of the wind may be heated to
temperatures $T\sim10^6$~K.

\end{abstract}

\section{Introduction}

High Mass X-ray Binaries (HMXBs) consist of a compact object (a neutron
star or black hole) accreting from an early type stellar companion.  
Stars of spectral types O and B typically drive stellar winds with mass
loss rates of $10^{-7}-10^{-5}\Msun$~yr$^{-1}$, and a portion of this
outflowing gas may be captured by the gravity of the compact object.
Simple analytic expressions of gravitational capture (e.g. Bondi \&\ Hoyle
1944) show that in some systems, the observed X-ray luminosity may be
accounted for entirely by accretion of the wind.  Other HMXBs, however, 
are probably powered by Roche lobe overflow leading 
to a gas stream and accretion disk.  Even systems in which it is 
thought that wind accretion 
dominates are often close to filling their Roche lobes (Petterson 1978).  
As the surface of the primary star approaches the critical potential
surface, there is probably a smooth transition between a stellar wind
enhanced along the line between the stars and Roche lobe overflow (Friend
\&\ Castor 1982).
  
One of the earliest tools for understanding these systems, suitable for
low resolution X-ray spectroscopy, was the examination of the orbital
variation of the X-ray absorption.  Buff \&\ McCray (1974) pointed out
that as an accreting compact object orbits a normal companion with a
spherically symmetric stellar wind, the wind should absorb soft X-rays at 
all phases of the orbit.
The amount of absorption should increase as the compact
object falls behind the densest parts of the wind, close to eclipse
ingress and egress. This technique has been used, for example, by Haberl,
White, \&\ Kallman (1989, hereafter HWK), who observed a binary orbit
of 4U~1700-37 using {\it EXOSAT}.  In this system, however, the absorption 
is not symmetric about orbital phase $\phi=0.5$ (Branduari et
al. 1978).  HWK showed that the excess absorption at $\phi>0.6$
could be explained by an accretion stream conveying 8\%\ of the
gas that accretes onto the compact object.

In addition to a gas stream, an ``accretion wake'', a bow shock trailing
the compact object, has also been invoked to explain X-ray absorption
in several systems (Livio, Shara, \&\ Shaviv 1979) and has been simulated
numerically (Blondin et al. 1990).

Another structure the compact object may create within the wind is a
``photoionization wake'', which results when the accelerating stellar wind
crashes into gas ionized by the X-ray source.  The ionized gas coasts, as
it lacks the ions that accelerate the wind through resonance scattering of
the stellar continuum (Fransson \&\ Fabian 1980). Photoionization wakes
are probably responsible for optical absorption lines from the primaries
in Vela~X-1 and 4U~1700-37 at $\phi>0.5$.  A gas stream or accretion wake
could cover the compact object but not enough of the primary to explain
the optical absorption (Kaper, Hammerschlag-Hensberge, \&\ Zuiderwijk,
1994).  Photoionization wakes appear in numerical simulations of X-ray
sources embedded in stellar winds for a wide range of X-ray luminosities
(Blondin 1994).

High resolution X-ray spectroscopy is emerging as a powerful diagnostic of
conditions in HMXBs.  An {\it ASCA} eclipse spectrum of the X-ray pulsar
Vela X-1 showed emission lines that could be attributed to hydrogen-like
and helium-like ionization stages of heavy elements (Si, Mg, and Ne, and
possibly S, Ar, and Fe) as well as neutral fluorescence lines. The line
strengths could be qualitatively understood through a model of
photoionization of the stellar wind (Sako et al. 1999).  {\it ASCA}
observations of the HMXB pulsar GX~301-2 also showed fluorescent lines
from Ne, Si, S, Ar, and Ca, as well as a 0.8~keV bremsstrahlung component
that may be caused by shocks trailing the neutron star (Saraswat et al.
1996).

The new generation of X-ray telescopes, {\it Chandra}-AXAF and {\it
Newton}-XMM, are confirming and expanding on the results suggested by {\it
ASCA} and are revealing diverse new phenomena.  Paerels et al. (2000)
observed emission lines and radiative recombination features in the HMXB
Cygnus~X-3 with {\it Chandra}.  Line ratio diagnostics suggest the
emission lines are formed by X-ray photoionization of the stellar wind.  
Schulz et al. (2002b) observed the Vela X-1 lines with much higher
resolution than with {\it ASCA} using {\it Chandra}, and resolved lines
that could be used to diagnose the state of the emitting gas.  The HMXB
black hole candidate Cygnus~X-1 shows variable emission and absorption
features (Schulz et al. 2002a), and may reveal the presence of a wind
focused by the gravity of the compact object (Miller et al. 2002).  The
superluminal jet source GRS~1915+105 shows absorption from H and He-like
Fe (Lee et al.  2002), while SS~433 shows relativistically Doppler-shifted
emission lines that provide density and temperature diagnostics for the
gas in the jets (Marshall, 
Canizares, \&\ Schulz, 2002).

A new dimension can be added to high resolution X-ray spectroscopy of
HMXBs when line features are observed through times when the X-ray
continuum is varying.  Power spectra of X-rays from HMXBs typically reveal
noise with a fractional rms variability of 20-30\%\ (Belloni \&\ Hasinger
1990).  Quasiperiodic oscillations (QPOs) with mHz frequencies have been
observed from X-ray pulsars such as LMC~X-4 (Moon \&\ Eikenberry 2001a)
and Hercules X-1 (Moon \&\ Eikenberry 2001b, Boroson et al. 2000).  
Observations of X-ray lines from a stellar wind with variable X-ray
illumination can test photoionization models.

For this investigation, we focus on the supergiant X-ray binary
4U~1700-37, discovered with {\it Uhuru} (Jones et al. 1973).
The X-ray source is eclipsed with a 3.41 day period 
by its companion HD~153919 (Penny et al. 1973, Hutchings et al. 1973),
determined to be an O6.5Iaf star of $M_O=52\pm2\Msun$ and $T_{\rm
eff}=42,000$~K. Doppler shifts of lines from HD~153919 indicate an orbit
that is nearly circular (Heap \&\ Corcoran 1992).  EXOSAT and
BATSE have set an upper limit of $\approx4$\%\ on coherent pulsations near
the expected $\sim100$~s period of a typical X-ray pulsar (Doll \&\
Brinkmann 1987, Rubin et al. 1996).  

Corbet (1986) showed that most HMXB X-ray pulsars are either orbiting Be
stars, or fall into separate period classes, 10$^2-10^3$~s and 1-10~s,
that may represent wind-fed and disk-fed systems, respectively.
The few reported detections of pulsations in 4U~1700-37 have not
been repeated.  Murakami et al. (1984) reported 67.4 second pulsations
during a flare observed with {\it Tenma}.  

There is no consensus on the cause of the flares in 4U~1700-37.
The X-ray light curve
shows periods, lasting on the order of an hour, when the flux increases by
up to a factor of $\sim10$.  These flaring periods are accompanied by
flickering in the light curve, also by about a factor of 10, on time
scales of seconds to about 10 minutes.
Brinkman (1981), based on a model of an entirely wind-accreting system,
suggested that the flares were associated with accretion from the
magnetotail of a neutron star.  Tentative detection of pulsations
suggested to Muakami et al. (1984) a similarity between 4U~1700-37 and
LMC~X-4, an X-ray pulsar that increases the fraction of pulsed X-rays
during flares in which the X-ray luminosity can reach several times the
Eddington luminosity.  Levine et al. (2000) suggested three possible 
causes for the flares in LMC~X-4:  an accretion instability, fusion on the
neutron star, or magnetic reconnection.  In contrast to LMC~X-4, in which
the spectrum becomes softer during flares, the spectral changes during
flares in 4U~1700-37 are not consistent, although there is evidence that
the soft X-ray excess can increase during flares (HWK).


The lack of a clear detection of pulsations in 4U~1700-37 is surprising,
as most neutron stars in HMXBs pulsate. The neutron stars in LMXBs are
rarely pulsars, but
they may have weaker magnetic fields owing to physical effects of
accretion over the longer life of the system (Bhattacharya \&\ Srinivasan
1995).

Measurements of the mass of the compact object in this system have been
contradictory. The mass can be determined from the mass function of the O
star,
\begin{equation}
f(M_O)\equiv P
K_O^3(1-e^2)^{3/2} /2\pi G=M_x\sin^3i/(1+1/q)^2 (1-e^2)^{3/2}
\end{equation}
where $K_O$ is the
orbital velocity of the O star, $P$ is the orbital period, $i$ is the
inclination, $q=M_x/M_O$ is the mass ratio, and $e$ is the eccentricity of
the orbit.  For $K_O$ in km~s$^{-1}$ and $P$ in days 
this gives:
\begin{equation}
f(M_O)=1.038\times10^{-7}K_O^3 P(1-e^2)^{3/2}
\end{equation}

Heap \&\ Corcoran (1992) using $K_O=18\pm3$~km~s$^{-1}$, as determined
from {\it IUE} observations, find $f(M_O)=0.0023\pm0.0005\Msun$, resulting
in an $M_x=1.8\pm0.4\Msun$. Stickland \&\ Lloyd (1993) re-examined the IUE
spectra that led to this determination, recalibrated the wavelength scale,
and found instead $K_O=5-10$~km~s$^{-1}$, leading conservatively to
$f(M_O)=0.00035\Msun$, providing an uncomfortably {\it low} mass for the
compact object.  Stickland and Lloyd raise the possibility that the
measured UV line variations do {\it not} give the radial orbital motion of
the O~star. Recently, Clark et al. (2002) using a measurement of
$K_O=20.6\pm1.0$ by Hammerschlag-Hensberge et al. (work not yet published
that involves reanalysis of spectroscopic data)  found
$M_x=2.44\pm0.27\Msun$, uncomfortably high for a neutron star but lower
than expected for a black hole candidate.

We list the measured parameters of the system in Table~1, including the 
range of mass determinations for the compact object is in a close orbit
($a\sin i=2.0\pm0.4R_O$, where $R_O=18\pm3$R$_\odot$, Heap \&\ Corcoran 
1992).

\section{Observations}

The {\it Chandra} observations we report here took place on August 22-23,
2000. We scheduled the observation to coincide with the period before
eclipse at $\phi=0.65-0.8$, so that we could investigate the pre-eclipse
absorption component and the soft excess. The {\it Chandra} observation
was coordinated with simultaneous observations with the Rossi X-ray Timing
Explorer ({\it RXTE}, Jahoda et al. 1996).

We used the High Energy Transmission Grating Spectrometer (HETGS,
Canizares et al. 2000) aboard the Chandra X-ray Observatory (Weisskopf et
al. 2002). We used the telescope's ``faint'' data mode, with a timed
readout every 3.24104 seconds. The HETGS includes both a Medium Energy
Grating (MEG) and a High Energy Grating (HEG) giving overall a spectral
range of 1-35\AA, with a resolution of $\lambda/\Delta \lambda=1400$ at 12
\AA\ and $\lambda/\Delta \lambda=180$ at 1.8 \AA.  The dispersed spectra
were then detected with the ACIS-S linear array of CCD chips.  The zeroth
order undispersed spectrum was severely piled-up, and was not used in our
analysis.

The observation was uninterrupted with a total exposure of
$4.241\times10^4$~s, including a deadtime correction of 1\%.  Including
photons in both the MEG and HEG, but excluding background, we detected
630,787 dispersed photons.

\subsection{Light Curves and Spectra}

We show a light curve for the combined HEG and MEG first order
spectra in Figure~1.  We have ignored photons near the wavelength of the 
Fe~K line, where pileup may flatten the light curve during flares.  

We have determined the
orbital ephemeris from RXTE ASM data to be $T_0=50088.625\pm0.030$ MJD, 
$P_0=3.411625\pm2\times10^{-5}$~days, and
$\dot{P}/P=-1.5\pm0.5\times10^{-6}$~yr$^{-1}$.

Along with the {\it Chandra} light curve, we also plot in red the 2--9~keV
light curve as observed with the Proportional Counter Array (PCA) on board
{\it RXTE}.  We show only counts from the PCU~3 unit, which is the most
reliable, although three PCUs were operating during most of the
observation. The agreement between the {\it Chandra} and {\it RXTE} light
curves is generally excellent, although the {\it RXTE} light curve appears
systematically brighter near flare peaks.  The {\it RXTE} observation
extended 41.4 hours before the {\it Chandra} observation and 6.8 hours
after.  The peak count rate in the {\it RXTE} PCU~3 over the entire {\it
RXTE} observation was about twice the peak count rate during the {\it
Chandra} observations.

For further analysis we divided the {\it Chandra} observation period into
three main segments (Figure~1), which we call Flaring (F), Quiescent (Q),
and Ending (E).  The Flaring period starts at the beginning of the
{\it Chandra} observation and ends at MJD 51778.6296, when the Quiescent
period begins.  (These times are spacecraft times, not corrected to the
solar system barycenter.)  The Quiescent period ends at MJD 51778.7454,
when the Ending period begins.

The division of the light curve into F, Q, and E periods is made
empirically based on the count rate.  The average Chandra count
rates in the (F,Q,E) periods are (6.4,0.7,1.7)~cts~s$^{-1}$.  However, as
we discuss in \S\ref{sec:analysis}, the absorbing column increases
throughout the observation, so that the intrinsic luminosity is more 
similar during the F and the E periods than the count rate would
indicate.  From simple broad-band fits of the spectra to absorbed power
law models (see \S\ref{sec:continuum}), we find average X-ray luminosities 
in the (F,Q,E) periods are
(1.1,0.2,0.8)$\times10^{36}$~erg~s$^{-1}$ (corrected for absorption).

In Figure~\ref{fig:spectra}, we show the 1-10\AA\ Chandra spectra of 
4U~1700-37 at each of the three periods of the observation.  
We have co-added HEG and MEG counts in the $\pm1$ 
orders.  The background is negligible and has not been subtracted.
We have applied standard CIAO procedures for ``destreaking'' the ACIS 
chips, and have verified that the strong lines do not overlap with chip 
gaps.  Although for display purposes we show the total counts in the 
$\pm1$ orders of the MEG and HEG added together, we fit the models to each 
order and instrument in parallel, using corresponding response files.
Fluorescence lines, notably an Fe~K line near 1.94\AA, are prominent in
both the F and E periods, and less prominent in the Q period.

High ionization lines due to H-like and He-like stages are also prominent
in the 4-12\AA\ wavelength range.
We identify these lines using House (1969), the analysis of lines in 
the Vela X-1 spectrum by Schulz et al. (2002b), and the XSTAR code
(Bautista 
\&\ Kallman 2001, Kallman \&\ Bautista 2001).

\section{Analysis\label{sec:analysis}}

To process and select counts, we used CIAO ({\it Chandra} 
Interactive Analysis of 
Observations) software version 2.2.1. 

We fit the continuum using the ISIS software package (Houck
\&Denicola 2000) and re-binning the
spectra by factors of 20 so that each bin has enough counts to allow use
of the $\chi^2$ statistic.  This analysis is described in 
\S\ref{sec:continuum}.

To fit the spectral lines, we used the Sherpa spectral fitting tool.  
Instead of $\chi^2$, we use
the C statistic (Cash 1979), which 
is appropriate when bins have few counts.  We analyze the fluorescence 
lines in \S\ref{sec:fluor} and the high ionization lines in 
\S\ref{sec:highion}.

For the spectral response files, we used {\tt
acisheg1D1999-07-22rmfN0004.fits} for the HEG and {\tt
acismeg1D1999-07-22rmfN0004.fits} for the MEG.  We created auxiliary
response files (ARFs) using the CIAO task ``fullgarf'' for event files for
each of the three periods of the observation (Flaring, Quiescent, and
Ending).

\subsection{Continuum Spectrum\label{sec:continuum}}

The hard continuum was fit using a power law absorbed by cold gas (a 
``Wisconsin 
absorber'', Morrison \&\ McCammon 1983).  We need a second component to 
fit the soft X-ray continuum.  We obtained adequate fits when we 
allowed for a second power law component with the same photon power law 
index but with a different normalization and behind a different column 
density.  Such a component could arise from Compton scattering in the 
wind.  Haberl \&\ Day (1992) fit {\it Ginga} spectra of 4U~1700-37 to a 
similar model, and Haberl (1991) performed Monte-Carlo simulations of 
scattering of the 4U~1700-37 spectrum in a stellar wind.

Parameters describing the continuum fits to the F, FH, FL, Q, E, and EL 
states are shown in Table~2.

The average spectrum during the Ending period with our
determination of continuum level and identification of spectral lines is 
shown in Figure~\ref{fig:lowenergy}.
 
\subsection{Fluorescent Lines\label{sec:fluor}}

We discuss the low-ionization fluorescence lines (Fe~K, Ca, Ar, S, Si, and
Mg, in order of increasing wavelength) observed with {\it Chandra}, paying
special attention to the brightest fluorescence feature, the Fe~K line 
near 1.94 \AA. In Figure~\ref{fig:fek} we show the spectrum in the region 
of the Fe~K lines during the F, Q, and E periods, along with our model 
fit.  The fit in Figure~\ref{fig:fek}a of the Fe~K lines during the Flaring 
period shows that the data are slightly in excess of the model.
We note that the model is {\it not} fit to the summed data, but that 
the spectrum in each HETG arm is fit in parallel.  The HEG~$+1$ and $-1$ 
spectra show Fe~K peaks offset by 0.005\AA\ from each other

The Fe~K line 
may be subject to some pileup during the Flaring period.  Two photons with 
energies $E_1$ and $E_2$ striking the detector within a frame time
register with an energy $E_1+E_2$.  While the number of piled up photons 
can in some cases be deduced from the ACIS-detected energy, the energy 
carried by piled-up Fe~K photons would exceed the ACIS threshold.  Pileup 
in the Fe~K line in the Flaring period would not only lead to an 
underestimate of the flux in the line, but also an overestimate of the 
line width.  

The edge apparent near 1.74\AA\ in Figure~\ref{fig:fek}c results from the
neutral Fe~K edge introduced by the ``Wisconsin'' model of interstellar
absorption we use to fit the continuum (wabs).

Fitting the Flaring period spectrum to an absorbed power law plus Gaussian
emission line gives a line centered at 1.937$\pm0.002$\AA\ (all errors are
90\%\ confidence and are entirely determined from fitting) with a Full
Width at Half Maximum (FWHM) of $2200\pm700$km~s$^{-1}$.  During the
Quiescent period, the centroid of the line is at a marginally longer
wavelength, 1.940$\pm0.002$\AA, with only an upper limit of
$2200$~km~s$^{-1}$ for the line width.  During the Ending period, the line
is centered on a wavelength of 1.936$\pm0.001$\AA\ but with a
significantly narrower FWHM of $900\pm400$~km~s$^{-1}$.

To identify the ionization stages of Fe responsible for the line, we use
the calculated K$\alpha$ line wavelengths of House (1969).  We find that
the 90\%\ confidence range is compatible with Fe~II--XII.

The most likely location for emission of the Fe~K line is the stellar 
wind.  HWK point out that the strength of the Fe~K line seen in 
mid-eclipse with EXOSAT is consistent with the out of eclipse flux
and the portion of the wind blocked by the normal star or not illuminated 
by X-rays during eclipse.  However, the possibility remains that some 
fluorescent emission arises on an accretion disk or gas stream.

Motion of the stellar wind can introduce wavelength changes via the
Doppler shift.  Based on high resolution UV spectra obtained with {\it
IUE}, Heap \&\ Corcoran estimate a wind terminal velocity of
$v_\infty=2100$~km~s$^{-1}$, while van Loon et al. (2001) estimate a
terminal velocity of $v_\infty=1700$~km~s$^{-1}$ and a fractional
microturbulent velocity of $\sigma=0.15$, which leads to a similar
Doppler broadening.  However, we are not able to separate the effects of
ionization stage, Doppler shift, and Compton scattering in the appearance
of the emission lines.

Comptonization can affect both the apparent wavelength of an emission line 
and its width (Kallman \&\ White 1989).  
Ignoring the weak temperature dependence in the Comptonization broadening,
the equations of Kallman \&\ White, based on models of Podnyakov, Sobol,
\&\ Sunyaev (1979), Ross, Weaver, \&\ McCray (1978), Langler, Ross, \&\
McCray (1978), and Illarionov et al. (1978), imply that $\Delta
\lambda/\lambda_0=0.0292 \tau_{\rm Th}$ where $\tau_{\rm Th}$ is the
optical depth for Thompson scattering.  
This implies $\tau_{\rm
Th}=0.25\pm0.08$ for the flaring period and $\tau_{\rm Th}=0.10\pm0.05$ 
for the ending period.  These values should be considered
upper limits, as some line broadening could also result from a range of
ionization stages in the emitting Fe or from Doppler shifts in the 
stellar wind, if the Fe~K line is emitted in the wind.

Assuming such a $\tau_{\rm Th}$, and that the line centroid wavelength
$\lambda$ is shifted by Comptonization from the rest wavelength
$\lambda_0$ according to $\lambda/\lambda_0=1+0.0375\tau_{\rm Th}$ (as
given by Kallman \&\ White, ignoring the temperature correction), we find
that our observations are compatible with all stages of Fe up to 
\ion{Fe}{18}.  We emphasize that these limits are set by the 
centroid of the Fe~K line and by its observed width.

The greatest scattering optical depth found from the line width,
$\tau_{\rm Th}=0.25\pm0.08$, found during the Flaring period, corresponds
to a column density $N_H=3.8\pm1.2\times10^{23}$~cm$^{-2}$.  The absorbing
column density seen throughout the orbit varies from
$\approx4\times10^{22}$~cm$^{-2}$ from $\phi=0.2$ to $\phi=0.6$, and
$2-4\times10^{23}$~cm$^{-2}$ from $\phi=0.8-0.9$ (HWK 1989).  Although 
EXOSAT did not observe from $\phi=0.6-0.75$, the models of HWK predict 
a range of column densities of $N_H=0.4-3\times10^{23}$~cm$^{-2}$.  The 
scattering optical depth $\tau_{\rm Th}$ decreases 
to $<1.5\pm0.8\times10^{23}$~cm$^{-2}$ in the Ending period.  Between the
Flaring and Ending periods, the Fe~K line is scattered by a {\it
decreasing} column density, in anticorrelation with the absorbing column
density, which {\it increases} between Flaring and Ending periods.
This could be the result of the particular scattering geometry involved.  
The absorbing column density samples the wind along the line of sight from 
the X-ray continuum source, whereas the Fe~K line is emitted in 
a more extended region in the wind (it persists through eclipse.)

During the F period, fits show only a marginal Fe~K$\beta$ line.  
However, the statistics are better during the E period, when the lines are 
narrower.  We fit the K$\beta$ line to a gaussian with central 
wavelength $\lambda=1.755\pm0.003$\AA and find a K$\beta$/K$\alpha$ ratio 
of $0.18\pm0.09$ (90\%\ confidence), consistent with the theoretical value 
of $\sim0.13$ for low optical depth in the lines (Kaastra \&\ Mewe 1993).

The 5.36\AA\ line which we attribute to 
S\,IV-VIII is absent in either of the low-luminosity states, as seen
in Figure~\ref{fig:3spectra}. 
There is marginal evidence for a line near 5.1\AA\ in the Quiescent 
spectrum that is not present in the Ending state.  This line could result 
from more highly ionized sulfur.

We also examine the relation between the continuum flux level and the
fluorescent line strength.  We show in Figures~\ref{fig:corel1}
and~\ref{fig:corel2} that overall the flux in the fluorescent lines
increase approximately linearly with the continuum flux.

\subsection{High Ionization Lines\label{sec:highion}}

The low energy (higher wavelength) lines, shown in 
Figure~\ref{fig:lowenergy}
are seen most prominently outside of the 
Flaring period, when the low-energy continuum has been diminished.
The lines are similar to those seen by Schulz et al. (2002b) during X-ray 
eclipse in the wind-fed 
X-ray binary Vela X-1.

The flux in the low energy lines is sufficient to account for the ``soft 
X-ray excess'', a feature of the X-ray spectrum previously without unique 
explanation.  HWK attributed this excess mainly to the passage of the X-rays 
through a partially ionized absorber, but they found that this model could 
not account for the excess in the spectrum at $\phi>0.6$.  HWK attributed 
the much larger excess seen at these phases to scattering.

The line widths and centroid wavelengths are given in
Table~\ref{tab:tablines}.  The fluxes given are the observed fluxes of the
lines, as they are behind an unknown column of absorbing gas.  The errors
in the line fluxes are probably overestimates, as they are based on the
errors in the FWHM and line amplitudes, and these errors may be
correlated.

These lines do not appear to be greatly variable as a result of 
either changes in orbital phase or the intermittent flares.

In Figure~\ref{fig:3spectra} we
compare the low energy lines seen in the Flaring, Ending and Quiescent 
periods.  We show only the spectrum during dips in the Flaring period, so 
that we can see the effect of the change in orbital phase.  We
have further subdivided the Ending period to separate periods in which the
count rate was similar to the count rate in the Quiescent period.  These
time intervals are indicated by horizontal bars in Figure~1. 


The low energy lines offer several diagnostics of the emitting gas.  The 
Helium-like triplets of \ion{Si}{13} and \ion{Mg}{11} are diagnostics of 
density and can be used to distinguish photoionization equilibrium from 
collisional ionization equilibrium (Porquet et al. 2002).  These triplets 
consist of a forbidden line (f, with line strength $z$), two closely spaced 
intercombination lines (i, with strengths $x$ and $y$), and a resonance 
line (r, with line strength $w$).  We show the He-like Si triplet during the E
period and the gaussian model in Figure~\ref{fig:helikesi}.

The fits to the \ion{Si}{13} lines (given in Table~\ref{tab:tablines}) give 
($w$,$x+y$,$z$)=$(10\pm3,3.8^{2.4}_{-1.8},,5.9^{+2.6}_{-2.1})$, where the 
errors are 
90\%\ confidence levels.
This result is incompatible with the assumption of a purely photoionized 
plasma, which would have $z>w$, and instead suggests a low density 
hybrid between photoionized and collisionally ionized gas. 

Anomalies in the He-like triplets in X-ray binaries are reported in 
Jimenez-Garate et al. (2002) and Wojdowski et al. (2002).  Jimenez-Garate et 
al. suggested from Newton-XMM observations of Hercules X-1 that the f levels 
could be excited to the i levels by UV photons.  However, this mechanism 
would not affect the r/f ratio.  In addition, the wavelength of 
the UV photon required to raise \ion{Si}{13} from the f to the i level is 
below the Lyman limit at 911\AA, and would be absorbed before reaching far 
in the wind.

Wojdowski et al. analyze Chandra 
observations of Centaurus~X-3 and find similar results to those presented 
here, namely a r/f ratio near unity, inconsistent with a purely photoionized 
gas.  Wojdowski et al. suggest that photoionization equilibrium still holds, 
but that resonance scattering of the continuum by the r lines adds to their 
flux during eclipse, when the direct continuum is diminished by absorption.
Outside of eclipse, however, one expects to see the r line superimposed with 
an absorption feature caused by scattering.

As can be seen in Figures~\ref{fig:lowenergy} and \ref{fig:3spectra}, the
ratio $r/f=w/z$ is $\gae1$ even in the low part of the Flaring period.
This is centered near $\phi=0.65$, far enough from eclipse that one might
expect to see diminution of the r lines from an absorption trough.  
In \S\ref{sec:lineform}, we discuss scattering scenarios in more detail.

The \ion{Mg}{11} triplet, seen in Figure~4b,c, is in the 
ratio $(w,x+y,z)=(6\pm3,3\pm2,4\pm2)$; unfortunately the error bars
are larger than for Si, but the ratio is still consistent with
a hybrid plasma.

\subsection{Line Formation Models\label{sec:lineform}}

We interpret the line strengths and ratios using the XSTAR photoionization 
code (Kallman \&\ Bautista 2001, Bautista \&\ Kallman 2001).

We used XSTAR version 2.1h, to compute the response of a gas to incident 
radiation.  The results depend most strongly on the ionization parameter 
$\xi\equiv L/n r^2$, where $L$ is the incident luminosity, $n$ is the 
particle density, and $r$ is the distance from the X-ray source, but also 
depend on the detailed source spectrum.  We used a power law spectral shape 
with a high energy cut-off as determined from BeppoSax observations 
(Reynolds et al. 1999).

Our line models were motivated by the suggestion, from the line ratios in
He-like Si, that the gas was not entirely photoionized but that dynamical
heating could play an important role as well.  Fransson \&\ Fabian for
example show that a photoionization wake can form in a wind-fed X-ray
binary, involving a collision of ionized gas and shadowed gas accelerated
by the normal stellar wind mechanism.  Temperatures $~10^6$~K can form
over about a stellar radius at an angle of 45 degrees from the X-ray
source.

Most low energy lines can be reproduced by a plasma fixed to
temperature $T=5\times10^6$~K and with an ionization parameter
$\log\xi=2$.  Such models do not predict any prominent RRC features,
consistent with the observations.  For high temperatures, RRC features are
broadened so that they are indistinguishable from continuum (although the 
absence of RRC features does not imply high temperature.)

However, concerns that temperatures reached in a photoionization wake
might not be high enough to contribute to the ionization of the X-ray
illuminated wind motivated us to study this problem further.  A pure
photoionization model could explain the anomalous He-like Si ratio through
resonance scattering. Resonance scattering is a zero-sum process, and one
might expect to have as many photons in the area scattered {\it away} as
scattered {\it toward} the viewer.  However, our point of view {\it is}
special in that the inclination of the orbit is high; at the orbital
phases of the {\it Chandra} observation, we may see the X-ray source
through an enhanced-density gas stream.  Our model of the continuum as
including a Compton-scattered component that passes through a lower column
density supports this.

If the gas is in pure photoionization equilibrium, it would have
implications for how the X-ray lines change with the ionizing continuum.

Assuming pure photoionization, we have computed a Differential Emission
Measure (DEM), a measure of the amount of gas at each bin of a parameter,
usually temperature for work on stellar coronae in thermal equilibrium or
ionization parameter $\xi$ in environments like gas around X-ray binaries.  
The problem as typically stated is overdetermined, that is, there are more
free parameters (for example, bins of $\log \xi$) than data points.  
However, a ``penalty function'' steers the solution toward one that is,
for example, as smooth as possible.  See Huenemoerder, Canizares, \&\
Schulz (2001) for a more detailed description of the method.  The major
change we have introduced is to use the ionization parameter $\xi$ instead
of the temperature $T$.

When we allow values of $\log \xi$ from $\log \xi=0$ to $\log \xi=5$, and
allow elemental abundances to be free parameters, we find that the DEM
during the Ending period has two peaks, one at $\log\xi<1$ (the
emission of the Fe~K line peaks at $\log\xi=1-2$) and the other at
$\log\xi\approx4$.

If the only difference between the Quiescent and Ending period is the
factor of $\sim4$ increased luminosity in the Ending period, one would
expect the DEM during the Q period to show a peak at $\log \xi=3.4$
instead of the $\log\xi=4$ peak in the E period.  Thus one would expect
emission lines formed at $\log\xi=2$ to be {\it brighter} during the Q
period than the E period.  As a result, we would expect the \ion{Si}{14}
line, which has peak emissivity at $\log\xi=3$, to be {\it dimmer} by a
factor of 2 during the Ending period than the Quiescent period, when in
fact we have observed the line to be 2.8$\pm0.7$ times brighter in the
Ending period.  The change in the fluorescent lines between the E and Q 
periods is, however, consistent with the DEM.  We find similar results 
when we freeze the elemental abundances to cosmic values.  In this case, 
the DEM predicts that the \ion{Si}{14} line would be 5 times brighter in 
the Q period.

Together with the anomalous He-like Si line ratios, this might suggest the
gas is not in pure photoionization equilibrium.  However, there are
large systematic uncertainties in the method and possible complications in 
the physical picture.  We suggest that if the anomalous He-like Si line 
ratio results from scattering, that further observations near $\phi=0.5$ 
might reveal an absorption trough instead of emission in the
\ion{Si}{13} resonance line.  Near $\phi=0.5$, the gas stream is out of
the line of sight to the X-ray source so that the 6-7\AA\ continuum we see
may be direct at not Compton scattered.  The contribution of a
resonance-scattered line component may be negligible against the direct
continuum.

\subsection{Power Spectra and Quasi-Periodic Oscillations}

We created power spectra of both the overall count rate and the counts 
at wavelengths $\lambda>4.5$\AA, displayed in Figure~\ref{fig:allpower} and 
Figure~\ref{fig:linepower}, respectively.  The power 
spectra are all Leahy-normalized (Leahy et al. 1983).

We fit the power spectrum of the integrated counts to an empirical model 
that consists of a power law with exponential cutoff, two Lorentzians, and 
a white noise component.  With Leahy normalization, the white noise from 
counting statistics would be fixed at 2, but detector dead time can reduce 
this slightly, so the white noise level is left as a free parameter.  We 
have binned the frequencies in the power spectra by a factor of 10 in 
order to determine errors in each bin.  The 
reduced $\chi^2$ value of our model of the power spectrum is 
$\chi^2_\nu=1.4$.

One Lorentzian component, detected with marginal significance, may
represent a quasiperiodic oscillation (QPO) with a centroid frequency of
6.5~mHz and a $Q=\nu/\Delta \nu$ value of $\approx15$.  This component
represents a fractional rms power of 1.4\%.  In fact, there are several
peaks in the 10$^{-3}$ to 10$^{-2}$~Hz range that may be QPO seen at low
significance.  Power spectral peaks with frequencies of exactly
$1,2\times10^{-3}$~Hz probably result from the spacecraft's 1,000 second
dither period.

The power spectrum of the region $\lambda>4$\AA\ shows more prominent QPO 
features (Figure~\ref{fig:linepower}).  We also allow a third Lorentzian 
at the second harmonic of the 6.5~mHz QPO.  The width of the second harmonic 
line is fixed to twice that of the fundamental.  The fit gives 
$\chi^2_\nu=2.1$, and it is clear that there are bumps in the 1--10 
mHz range that the model does not account for.  The 6.5~mHz line has a 
fractional rms amplitude of 4.5\%\ while its second harmonic has an 
amplitude of 4.0\%.  

Similar temporal behavior has been seen from 4U~1700-37 before in the 
integrated counts 
observed with EXOSAT (Doll \&\ Brinkmann 1987).  Doll \&\ Brinkmann argued 
that the timescales of variability in 4U~1700-37 reflected the flare 
recurrence rate and was linked to wind flow times.  

Examination of the cross-correlation coefficient between the hard and soft 
X-rays shows a peak consistent with zero delay between the two.

We have also formed a power spectrum without any binning, in 
order to see 
more clearly the low frequency variations in the $\lambda>4$\AA\ spectrum.  
This power spectrum is 
displayed in Figure~\ref{fig:fullpsd}.  To the eye there appears to be a 
cutoff near the frequency $\nu=3\times10^{-4}$~Hz.
The time scales of the 
cutoff and QPOs (10$^2-10^4$ seconds) lie between the light travel-time and 
wind-travel time between the stars.  Our continuum models suggest that 
much of the flux in this wavelength range results from Compton scattering.  
The cutoff in the power spectrum could then result from smearing of 
the X-ray variability by Compton scattering.

Although the region $\lambda>4$\AA\ is dominated by lines during the 
Ending period (Figure~\ref{fig:3spectra}), it is dominated by the continuum 
during the Flaring period (Figure~\ref{fig:spectra}).  Thus the low 
frequency noise and stronger QPOs in the $\lambda>4$~\AA\ region probably 
result from the continuum and not the lines.  However, restricting analysis 
to only the wavelength regions surrounding the lines and to only the 
Quiescent and Ending periods, we still find significant low frequency noise.

\section{Conclusions and Future Work}

The current data set and the 4U 1700-37 system offer outstanding diagnostics 
of stellar winds and their disruption by compact objects.  

The low energy lines (excluding the S line at 5.36\AA\ and the Si line at
7.11\AA) arise in highly ionized gas, yet a soft component is still
visible through eclipse (HWK).  XSTAR simulations can reproduce the low
energy lines either in photoionization equilibrium with $\log\xi=2.5-3.0$
or as a hybrid plasma at $T=5\times10^6$~K with $\log\xi=1.5-2.0$ (at the
lower end of that range the \ion{Si}{14}$\lambda6.18$ to
\ion{Si}{13}$\lambda\lambda\lambda$6.64,6.68,6.74 ratio becomes lower than
observed). In Figure~\ref{fig:xi} we show that for a uniform wind model, a
region in photoionization equilibrium with $\log \xi=2.5$ is smaller than
the primary and would not be visible in mid-eclipse. While this would seem
to support an extended region of hot gas (perhaps a photoionization wake)
real stellar winds are probably highly nonuniform, with density contrasts
of $\sim1000$.  In such a nonuniform wind, there could be pockets of gas
with $\log\xi=2.5$ that extend all the way to the region indicated by
$\log\xi=0$ for the uniform wind case.

The Fe lines (and the low ionization S$\lambda5.36$ and Si$\lambda7.11$ 
lines) may arise in a region with $\log\xi\lae1$.  This region could be 
large enough for the lines to be visible through mid-eclipse even for a 
uniform wind.  

Follow-up observations can test whether there is a variable delay between 
the hard and soft X-ray oscillations.  This could result, for example, if 
the soft X-rays arise in a more extended region in response to hard X-rays 
from the compact object.  During the present observations, we detected no 
delay.  

QPO with mHz frequencies have been seen from a number of X-ray binary
pulsars (Boroson et al. 2000).  The detection of mHz QPOs in 4U~1700-37
thus suggests that the compact object is a neutron star.  Further
observations, detecting QPOs at several X-ray luminosity states, can can
test any relationship between the frequency and an inner-disk radius.

The cause of the anomalous ratio in the \ion{Si}{13} triplet has not been
settled, but the two possible explanations, resonance scattering or a
hybrid plasma, would both have important implications.  The optical depth
at 6.7\AA, and thus density in the wind would be constrained by the
requirement that enough photons could be scattered.  On the other hand, if
the line ratios are caused by hot gas, this contributes evidence for and
helps diagnose the physical extent and state of a photoionization wake.

One promising method for diagnosing resonance scattering is high
resolution X-ray spectroscopy, but near $\phi=0.5$, as described in
\S\ref{sec:lineform}.

If there is extended hot gas, we would expect it to contribute to the
continuum emission. Such a component has been used to fit the continuum
spectrum before. Haberl \&\ Day (1992) used a bremsstrahlung component
with $kT=0.5$~keV behind a column density N$_H=5\times10^{21}$~cm$^{-2}$
to fit {\it GINGA} spectra (as well as the {\it EXOSAT} data of HWK).  
ROSAT observations found a bremsstrahlung component with $kT=0.47$~keV and
$kT=0.74$~keV after eclipse (Haberl, Aoki, \&\ Mavromatakis, F., 1994).  
Most recently, BeppoSax observations found a $T=0.2\pm0.1$~keV component 
(Reynolds et al. 1999). After adjusting the normalization of the HEG and
MEG spectra, we find that we do not require a bremsstrahlung component to
fit the broad-band spectrum during our {\it Chandra} observations.  
However, our models of the broad-band continuum are hampered by
uncertainties in the HEG and MEG effective areas, and a bremsstrahlung
component may be present at flux levels lower than historically observed.

Other HMXB, such as Cen X-3 (Wojdowski et al. 2003) or Vela X-1 (Schulz et 
al. 2002) also show similar \ion{Si}{13} diagnostics, but these have been 
interpreted to be result of resonance scattering.  Given that Vela~X-1 
probably also has a photoionization wake (Kaper et al. 1994), emission 
from a hybrid plasma should also be considered as a possibility, 
particularly for that system.  Again, analysis of observations at a 
variety of orbital phases can help distinguish between the possibilities.  
Hydrodynamic models of individual systems can also suggest whether the 
wind can be heated sufficiently by dynamical means.

\acknowledgements

SDV supported in part by NASA
(NAG5-6711), and the Chandra X-ray Center (GO0-1100X).  We would like to 
thank Mr. Andrew Beltz and Mr. Corey Casto for their assistance.  We would 
like to thank the referee for comments.

\clearpage

\begin{deluxetable}{lccc}
\label{tab:parameters}
\tablewidth{0pt}
\tablecaption{Parameters of the 4U 1700-37 System}
\tablehead{
\colhead{Parameter name} & \colhead{Expression} & \colhead{Value} &
\colhead{Reference\tablenotemark{a}}
}
\startdata
\tableline
Distance & $D$ & 1.9 kpc & 1\\
Temperature of O star &  T$_*$ & 42000$\pm2000$, 35000$\pm1000$~K & 2, 4\\
O Star Radius & R$_O$ & 18$\pm3$, 21.9$^{+1.3}_{-0.5}$R$_\odot$ & 2, 4\\
Mass of O star & M$_O$ & 52$\pm2$,58$\pm11$~M$_\odot$ & 2, 4\\
Mass of compact star & M$_x$ & 1.8$\pm0.4$, 2.44$\pm0.27$~M$\odot$ & 2, 4\\
Spectral Type & & O6.5Iaf & 2\\
Orbital inclination & $i$ & $>80^\circ$ & 2\\
Orbital separation & $a \sin i$ & $2.0\pm0.4$ R$_O$ & 2\\
Wind terminal velocity & v$_\infty$ & 2100$\pm400$, 1700 & 2,3\\
Wind mass loss rate & $\dot{M}$ & 6, 9.5$\times10^{-6}\Msun$~yr$^{-1}$ & 
2, 4\\
\enddata

\tablenotetext{a}{1: Ankay et al. 2001, 2: Heap \& Corcoran 1992, 
3: van Loon et al. 2001, 4: Clark et al. 2002}
\end{deluxetable}

\begin{deluxetable}{lcccccc}
\tablewidth{0pt}
\tabletypesize{\scriptsize}
\tablecaption{Fits to the X-ray Continuum}
\tablehead{
\colhead{Parameter} & \colhead{F\tablenotemark{a}} & 
\colhead{FH} & \colhead{FL} & 
\colhead{Q}
& \colhead{E} & \colhead{EL}}

\startdata


Exposure time (s)\tablenotemark{b} & 11420 & 6164 & 5256 & 10000 & 21531 
& 
3300\\

N$_H$ (10$^{23}$~cm$^{-2}$) &  
0.91$^{+0.14}_{-0.07}$ & 
0.92$^{+0.19}_{-0.12}$ &
1.06$^{+0.11}_{-0.06}$ & 
1.27$\pm0.03$ & 
2.03$^\pm0.08$ & 
1.8$^{+0.3}_{-0.1}$\\

Power Law Norm\tablenotemark{c} &
 0.16$_{-0.01}^{+0.03}$ & 
0.17$^{+0.04}_{-0.02}$ & 
0.18$^{+0.06}_{-0.03}$  
& 0.038$^{+0.020}_{-0.010}$ & 
0.095$\pm0.018$ & 
0.04$^{+0.03}_{-0.02}$\\

Power Law $\gamma$\tablenotemark{d} & 
1.09$_{-0.07}^{+0.10}$ & 
1.04$^{+0.13}_{-0.08}$ & 
1.30$^{+0.15}_{-0.11}$ &
1.30$^{+0.24}_{-0.13}$ 
& 1.00$^\pm0.10$
& 1.2$^{+0.4}_{-0.3}$\\

N$_H$ (2) (10$^{22}$~cm$^{-2}$) &
 3.4$\pm0.5$ & 
3.6$^{+0.5}_{-0.4}$ &
2.0$\pm1.5$ &
0.29, $<3.9$ & 
0.95 ($<1.8$) &
0.97 ($<2.1$)\\

Power Law (2) Norm  &
 0.03$^{+0.02}_{-0.01}$  & 
0.06$^{+0.04}_{-0.02}$ & 
0.002$^{+0.006}_{-0.001}$ &
$1.0\pm0.4\times10^{-4}$ &
3$^{+2}_{-1}\times10^{-3}$ & 
4$\pm2\times10^{-3}$\\

$\chi^2_\nu, \nu$ & 1.06,607 & 1.15,301 & 1.13,357 & 1.09,61 & 1.07,301 &
(c-stat\tablenotemark{d}, 61)

\enddata
\tablenotetext{a}{F=Flaring, FH=Flaring High, FL=Flaring Low, Q=Quiescent, 
E=Ending, EL=Ending Low, as shown in Figure~1}
\tablenotetext{b}{The total on-source time, including non-readout time 
(``dead time'')}
\tablenotetext{c}{The normalization is defined as the photons keV$^{-1}$ 
cm$^{-2}$ s$^{-1}$ at 1~keV.}
\tablenotetext{d}{This is the ``photon index'', so that the number of 
photons at energy $E$ is given by $K (E/1\mbox{keV})^{-\gamma}$, where $K$ 
is the normalization}
\tablenotetext{e}{For the low ending state, we used the c-statistic (Cash 
1979) instead of $\chi^2$, as even when the wavelength bins were grouped, 
the statistics were too poor to apply $\chi^2$.} 

\end{deluxetable}

\begin{deluxetable}{lccccc}
\tabletypesize{\scriptsize}
\label{tab:tablines}
\tablewidth{0pt}
\tablecaption{Measurements of Line Wavelengths, Widths, and Fluxes}
\tablehead{
\colhead{$\lambda$ (\AA)} & \colhead{Mode\tablenotemark{a}} & 
\colhead{Ion} & 
\colhead{$\lambda_0$ (\AA)\tablenotemark{b}} & 
\colhead{FWHM (km/s)} & \colhead{Flux (10$^{-6}$ photons s$^{-1}$
cm$^{-2}$)}}
\startdata
\tableline

1.753$\pm0.001$ & F & Fe\,II-XII,K$\beta$\tablenotemark{c} & 1.756 & 
2200$\pm700$ & 
300$\pm200$\\

1.76$\pm0.01$ & FH & & & 2000$\pm900$ & 300$\pm300$\\

1.75$\pm0.01$ & FL & & & 2000$\pm900$ & 200$\pm200$\\

1.755$\pm0.003$ & E & & & 900$\pm400$ & 200$\pm100$\\

1.937$\pm0.002$ & F & Fe\,II-XII,K$\alpha$ & 1.936-1.937 &
2200$\pm700$ & 1200$\pm200$ \\

1.937$\pm0.002$ & FH & & & 2000$\pm900$ & 1400$\pm300$\\

1.937$\pm0.002$ & FL & & & 2000$\pm900$ & 900$\pm200$\\

1.940$\pm0.002$ & Q & & & $<2200$ & 170$\pm60$\\

1.936$\pm0.001$ & E & & & 900$\pm400$ & 1100$\pm90$\\

1.938$\pm0.001$ & EL & & & unresolved & 350$\pm130$\\

3.355$\pm0.001$ & E & Ca\,VI-VII & 3.356-3.354 & unresolved & 46$\pm15$\\

4.18\tablenotemark{d} & F & Ar VI-IX & 4.186-4.178 & unresolved &
$40\pm30$\\

4.18\tablenotemark{d} & FL & & & unresolved & 66$\pm40$\\

4.18\tablenotemark{d} & Q & & & unresolved & $<13$\\

4.18\tablenotemark{d} & E & Ar VI-IX & 4.186-4.178 & unresolved & 
20$\pm9$\\

4.18\tablenotemark{d} & EL & & & unresolved & 30$^{+30}_{-20}$\\

4.744$\pm0.004$ & F & S\,XVI & 4.7274-4.7328 & 
unresolved & 
60$\pm30$\\

4.744$\pm0.002$ & FH & & & unresolved & 90$\pm50$\\

4.69$^{+0.05}_{-0.02}$ & FL & & & unresolved & 
30$\pm30$\\

4.74\tablenotemark{d} & E   &  &  & unresolved &  $14\pm7$ \\

4.74\tablenotemark{d} & EL & & & unresolved & $<22$\\

5.10$\pm0.01$ & Q & S\,XIII-XV & 5.075-5.161 & undetermined & 
14$^{+12}_{-9}$\\

5.36\tablenotemark{d} & F & S\,IV-VIII & 5.370-5.356 & $<3000$ & 50$\pm30$\\

5.366$\pm0.005$ & FH &  &  & $<4000$ & 80$\pm50$\\

5.36\tablenotemark{d} & FL & & & unresolved & 40$^{+40}_{-30}$\\

5.37$\pm0.03$ & Q & & & $>1000$ & 12$\pm12$\\

5.359$\pm0.005$ & E &  &  & 3000$\pm1000$ & 70$\pm15$\\

5.36$\pm0.02$ & EL & & & 3000 & 30$^{+30}_{-20}$\\

6.17 & F & Si\,XIV & 6.199 & unresolved & 
29$\pm10$\\

6.17 & FH & & & unresolved & 30$\pm16$\\

6.22$\pm0.01$ & F & ? & & 10000 & 30$\pm14$\\

6.22$\pm0.01$ & FH & ? & & 10000 & 47$\pm25$\\

6.17 & FL &  & & unresolved & 27$\pm11$\\

6.18$\pm0.01$ & Q & & & 1800$^{+1800}_{-700}$ & 6$^{+4}_{-3}$\\

6.177$\pm0.003$ & E &  &  & 1200$\pm300$ & 17$\pm4$\\

6.179$\pm0.001$ & EL & & & 1200 & 13$^{+10}_{-7}$\\

6.62$\pm0.01$ & F & Si\,XIII & 6.647 & 
1900$\pm500$ & 38$\pm12$\\

6.61$\pm0.02$ & FH & & & 1200$\pm500$ & 39$\pm19$\\

6.633$\pm0.006$ & FL & & & 1000$\pm500$ & 25$\pm10$\\

6.63$\pm0.01$ & Q & & & 2000$\pm1000$ & 11$^{+5}_{-4}$\\

6.640$\pm0.005$ & E   &  &  & $1200\pm400$ & 
$10\pm3$\\

6.640\tablenotemark{d} & EL & & & 1200 & 10$^{+9}_{-6}$\\

6.68$^{+0.07}_{-0.03}$ & Q & Si\,XIII & 6.684,6.687 & 
1400$^{+1500}_{-600}$ & 1.9$^{+3.0}_{-1.9}$\\ 

6.684$\pm0.009$ & E &  & & 	1200$\pm400$\tablenotemark{e}	
& $3.8^{+2.4}_{-1.8}$\\

6.684\tablenotemark{d} & EL & & & 1200 & $<3$\\

6.73$\pm0.01$ & F & Si\,XIII & 6.739 & 1900$\pm500$ & 33$\pm12$\\

6.73$\pm0.01$ & FH & & & 1200$\pm500$ & $43\pm20$\\

6.73$\pm0.01$ & FL & & & 1000$\pm500$ & 11$^{+9}_{-7}$\\

6.73$\pm0.02$ & Q  & & & 2500$^{+1500}_{-900}$ & 6$^{+4}_{-3}$\\

6.733$\pm0.007$ & E & Si\,XIII & 6.739 & 1200$\pm200$ & 
$5.9^{+2.6}_{-2.1}$\\

6.733\tablenotemark{d} & EL & & & 1200 & $<8$\\

7.05$\pm0.06$ & F & & & 13000$^{+10000}_{-4000}$ & 50$\pm20$\\

7.00$\pm0.11$ & FH & Si blend & & 8000$^{+5000}_{-3000}$ & 120$\pm40$\\

7.09$\pm0.04$ & FL & & & 8000$^{+5000}_{-3000}$ & 41$\pm15$\\

7.02$\pm0.03$ & E & Si blend &  & 8000$^{+3000}_{-2000}$ & $12\pm4$\\

7.099$\pm0.004$ & F & $<$Si\,VII & $>7.063$ & $1100\pm400$ & 40$\pm10$\\

7.099$\pm0.007$ & FH & &  & 1100$^{+700}_{-400}$ & 52$\pm16$\\

7.098$\pm0.002$ & FL & & & 800$\pm400$ & 32$\pm10$\\

7.110$\pm0.007$ & Q & & & $1700^{+1200}_{-700}$ & 14$\pm5$\\

7.110$\pm0.003$ & E &  &  & 1300$\pm300$ & $22\pm4$\\

7.110$\pm0.01$ & EL & & & 1300 & 13$\pm7$\\

7.84$\pm0.02$ & E & Mg\,XI & 7.850 & 2100$^{+1800}_{-700}$ & 3$\pm2$\\

8.41\tablenotemark{f} & FH & Mg\,XII & 8.419,8.424 & $<1400$ & 14$\pm7$\\
  & FL & & & 1100$\pm500$ & 13$\pm6$\\ 
  & Q & & & $<8000$ & 6$\pm3$\\
  & E & & & 1300$\pm700$ & $5\pm2$\\
9.1681 & FH & Mg\,XI & & 3000$^{+3000}_{-1000}$ & 21$\pm10$\\ 
       &  E &    & & 2500$^{+2000}_{-800}$ & $4\pm2$\\
9.2280 & FH & Mg\,XI & & 3000$^{+3000}_{-1000}$ & $<10$\\
        & E &        & & 2500$^{+2000}_{-800}$ & $3\pm2$\\
9.3134 & FH & Mg\,XI & & 3000$^{+3000}_{-1000}$ & $<14$\\
       & E &          & & 2500$^{+2000}_{-800}$ & $6\pm3$\\
10.238 & FH & Ne\,X\,Ly$\beta$ & & unresolved & 7$^{+8}_{-5}$\\
  & E &  & & unresolved  & 3$\pm2$	\\
12.132 & FH & Ne\,X\,Ly$\alpha$ & & unresolved & 8$^{+10}_{-6}$\\
       &  E &    & & unresolved & $4\pm3$\\
\enddata
\tablenotetext{a}{F=Flaring, FH=Flaring High, FL=Flaring Low, Q=Quiescent, 
E=Ending, EL=Ending Low, as shown in Figure~1}
\tablenotetext{b}{Rest wavelength.}
\tablenotetext{c}{We give wavelengths for K$\alpha_1$.
The centroids are $\approx0.002$\AA\ greater than the $\alpha_1$ 
wavelengths.  The fluxes we give are for 
K$\alpha_1$+K$\alpha_2=1.5$~K$_{\alpha_1}$}
\tablenotetext{d}{Marginal detection.  The wavelength was frozen to the
given value}
\tablenotetext{e}{The He-like Si lines are fixed to have equal FWHM.}
\tablenotetext{f}{All the lines with $\lambda>8$\AA\ have their 
wavelengths frozen to the given value}
\end{deluxetable}

\clearpage

\figcaption{The combined MEG and HEG light curve of 4U 1700-37.  
We have divided the light curve into three separate intervals, 
which we call Flaring, Quiescent, and the Ending periods.  To separate the 
effects of orbital phase and X-ray flaring, we investigate times in the F 
period when the X-ray light curve, smoothed over 30 pixels, is less than 
30 cts per 3.24 second readout bin.  We also 
 investigate the spectrum during times in the E period in which the count
rate resembles the count rate during the Q period.  The separate intervals
in the F and E periods that are selected for further analysis are
indicated by the horizontal lines with bars at the ends.  We overplot in
red the contemporaneous RXTE light curve.  The scale of the RXTE light
curve is given on the right $y$ axis.}

\figcaption{The combined MEG and HEG spectrum of 4U 1700-37 integrated 
over the Flaring period (F), Quiescent period (Q), and Ending period (E) 
of our observation.  The 1.93\AA\ Fe K line is apparent, as well as lines 
in the 4-8\AA\ range in the F and E spectra.  The spectra have been 
binned to 0.04\AA\ resolution.  We also show our model of the continuum, 
which consists of a power law behind two different column densities.  The 
softer, less-absorbed spectrum is plotted with dashed lines, while the 
harder, more-absorbed component is plotted with dotted lines.  The 
colors are chosen consistently within each observing segment (F,Q,E).}

\figcaption{The Fe K$\alpha$ and K$\beta$ lines during the (a) Flaring 
(0.005~\AA/pixel), 
(b) Quiescent (0.01~\AA/pixel), and 
(c) Ending (0.005~\AA/pixel binning) periods.  The solid curve is the 
model fit, while the histogram 
is the combined MEG and HEG data.  The edge near 1.74\AA\ results from 
neutral iron and is implicit in the model for cold absorption that the 
continuum (power law) passes through.}

\figcaption{The 1.5-13\AA\ spectrum during the Ending period, with 
Gaussian fits to the lines and line identifications.  We show 
(a) the count spectrum, (b) the log of the count spectrum, (c) the 
spectrum with the continuum model subtracted.}

\figcaption{A comparison of the MEG spectra (in counts per second per 0.02
\AA\ bin) at different times.  From top to bottom, the panels show: the
spectrum during the low count rate intervals of the Flaring period, the
spectrum integrated over the entire Quiescent period, the spectrum
integrated over the entire Ending period, and the spectrum during the low
count rate intervals of the Ending period. The error bars show the 
1$\sigma$ errors for the 7.1\AA\ line peak.  The times of the low count rate
periods are indicated by the horizontal bars in Figure~1.}

\figcaption{Correlations between the flux in the X-ray continuum (1-13AA) and
the flux measured in fluorescent lines.  For the continuum flux, we include
both the direct and Compton-scattered components at the periods (FH,FL,Q,E,EL).
a),b),c),d), and e) denote whether the fluorescence feature is due to Fe~K, Ar, S,
narrow Si near 7\AA, or broad Si near 7\AA.  All error bars are 90\%}

\figcaption{This is identical to the previous plot, but now the continuum includes
only the hard component but without any absorption.  Again,
the fluorescence feature is caused by a) Fe~K, b)  Ar, c) S,
d) narrow Si near 7\AA, or e) broad Si near 7\AA.}

\figcaption{The spectrum near the He-like Si-triplet in the Ending period.  The
rest wavelengths of the Resonance, two Intercombination lines, and the
Forbidden lines are shown with vertical dotted lines.}

\figcaption{The Leahy-normalized power spectrum of the continuum.  We have 
subtracted 
counts in the low energy lines from the total counts in HEG and MEG 1st 
order.  The total model and individual model components are also shown.  
We show error bars where the logarithmic scale allows them to be visible.}

\figcaption{The Leahy-normalized power spectrum of the low energy region.  
The total model and individual model components are also shown.}

\figcaption{The Leahy-normalized power spectrum.  In order to emphasize 
the low-frequency behavior, we have not performed any binning.}

\figcaption{Contours of constant ionization parameter $\log\xi$ for a simple 
model of a uniform wind with $\dot{M}=6\times10^{-6}\Msun$~yr$^{-1}$, a 
terminal velocity of $v_\infty=1700$~km/s, a wind velocity law of 
$v(R)=v_\infty(1-R_O/R)^\beta$ where $\beta=0.8$.  The ``x'' marks the 
position of the compact object and the circle shows the primary.  The 
$\log\xi=0.0$ ionization boundary is cut off where the primary's shadow 
prevents X-rays from reaching the wind.}

\clearpage

\setcounter{figure}{0}

\begin{figure}
\caption{}
\plotone{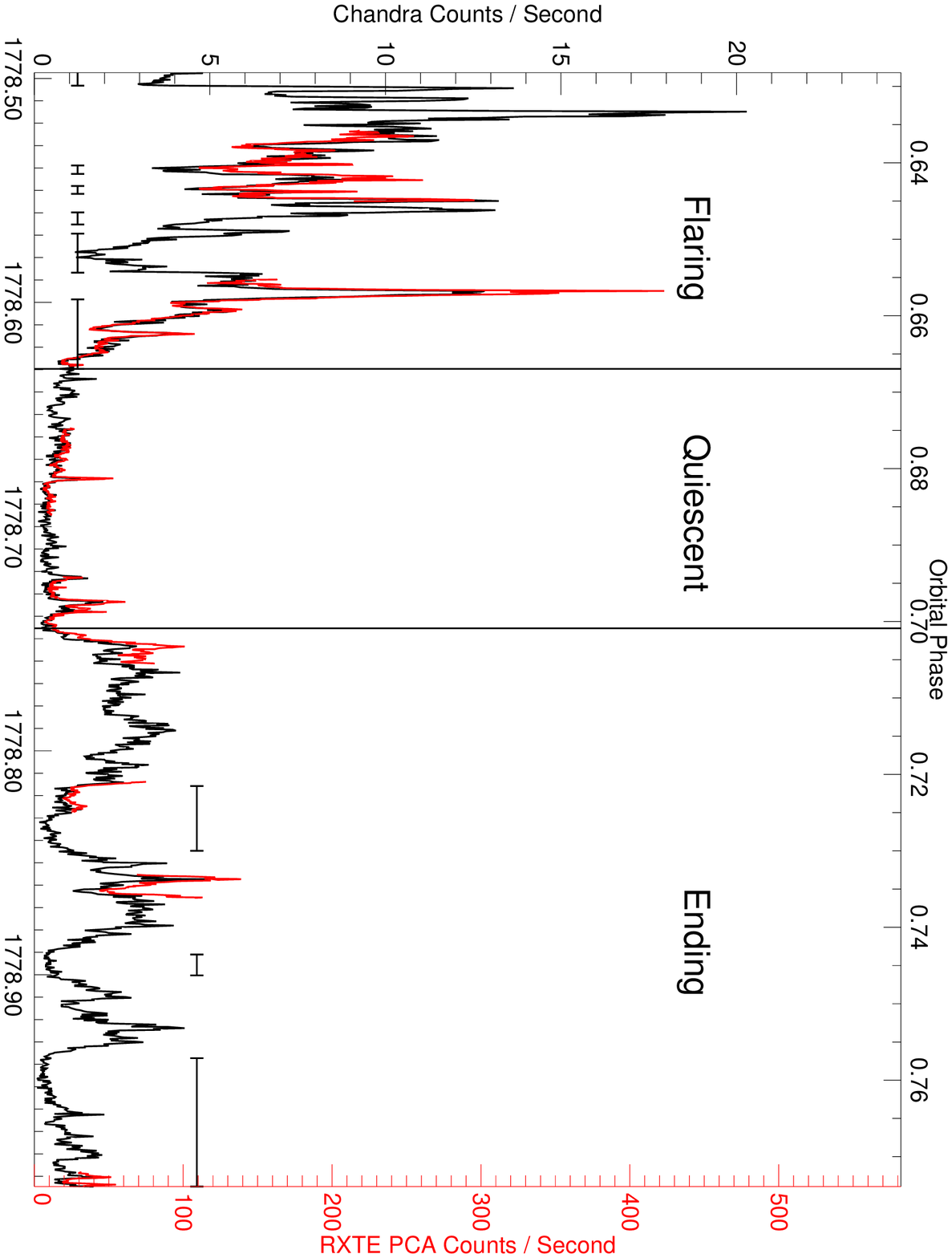}
\end{figure}

\begin{figure}
\caption{
\label{fig:spectra}
}
\plotone{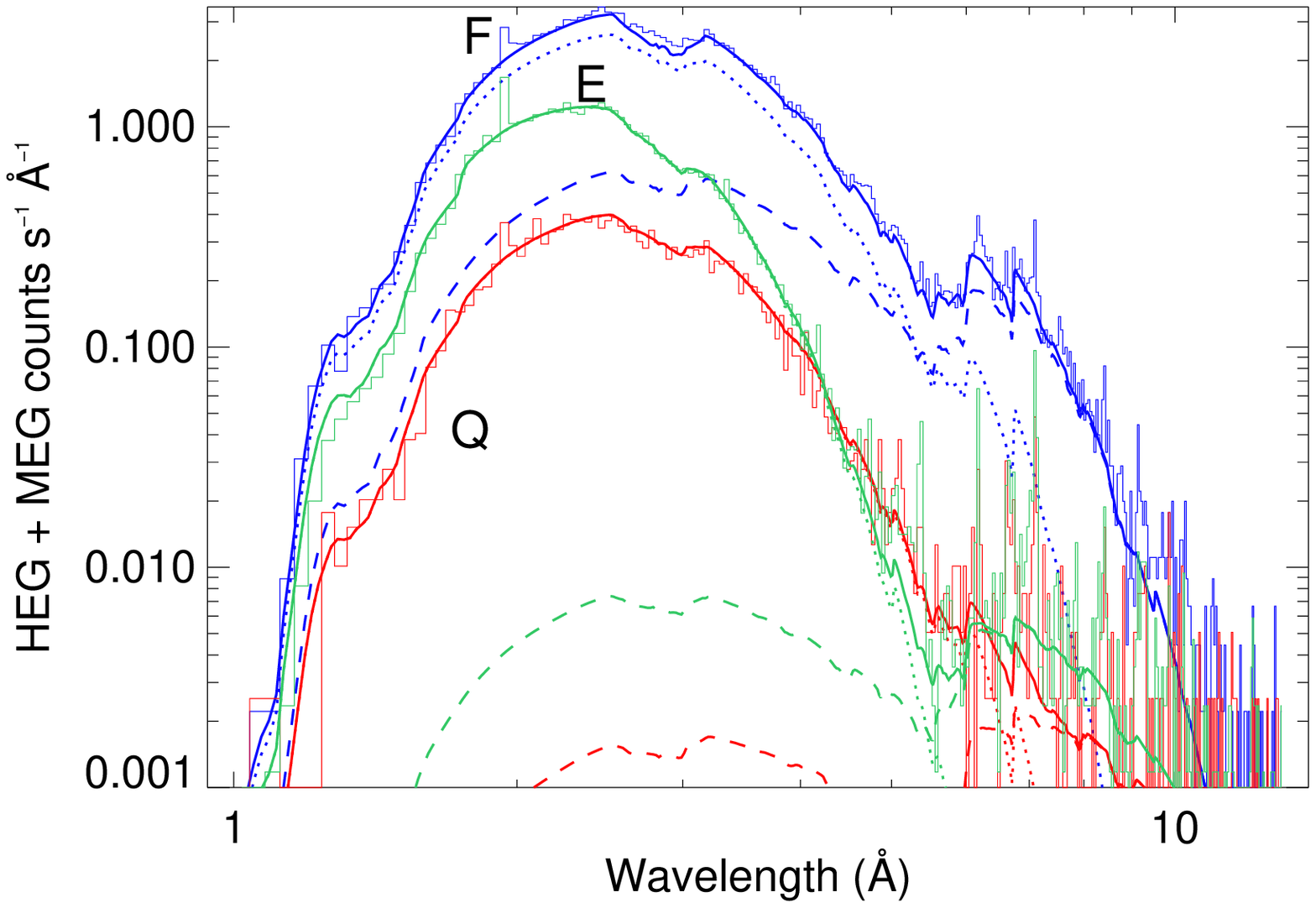}
\end{figure}

\begin{figure}
\caption{\label{fig:fek}
}
\plotone{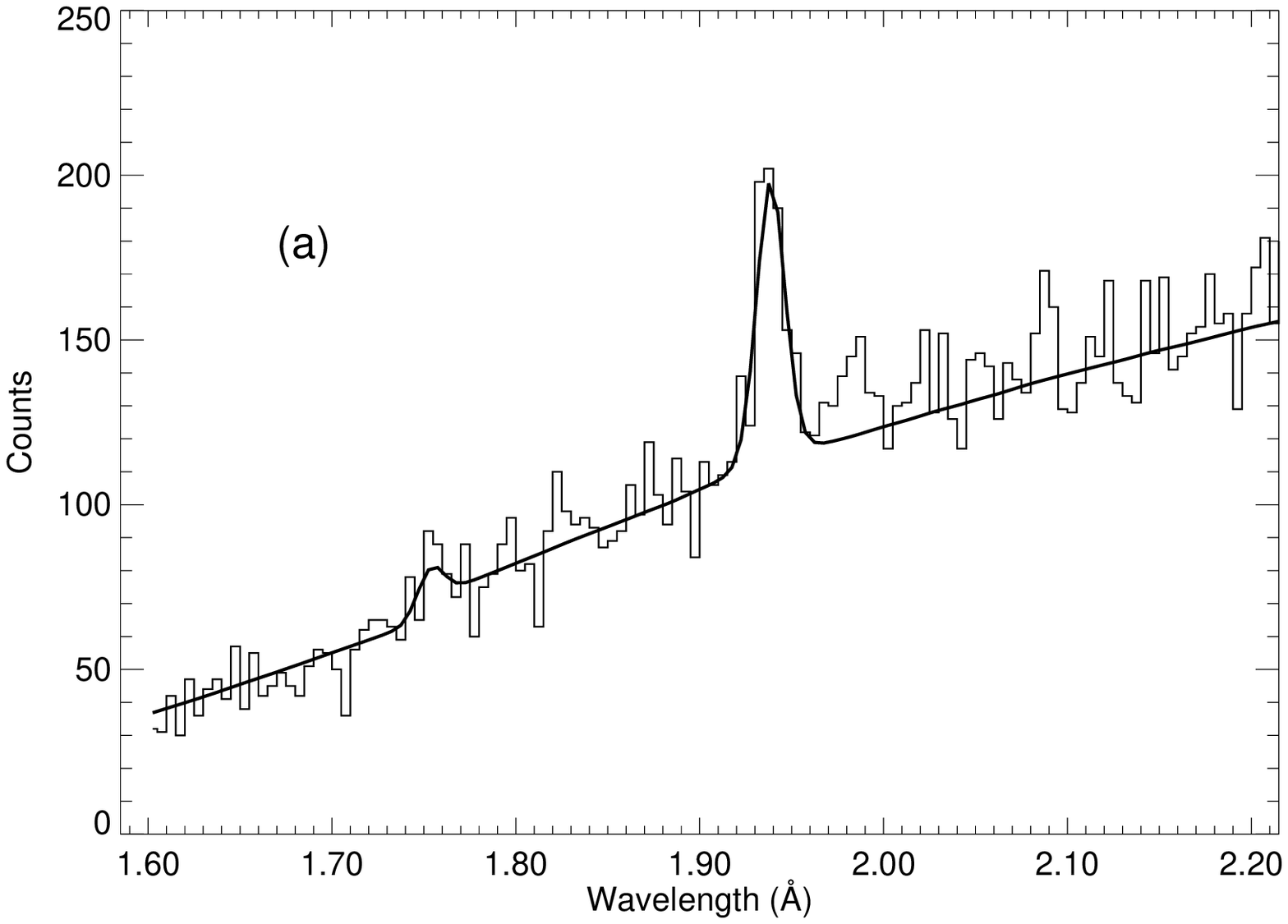}
\plotone{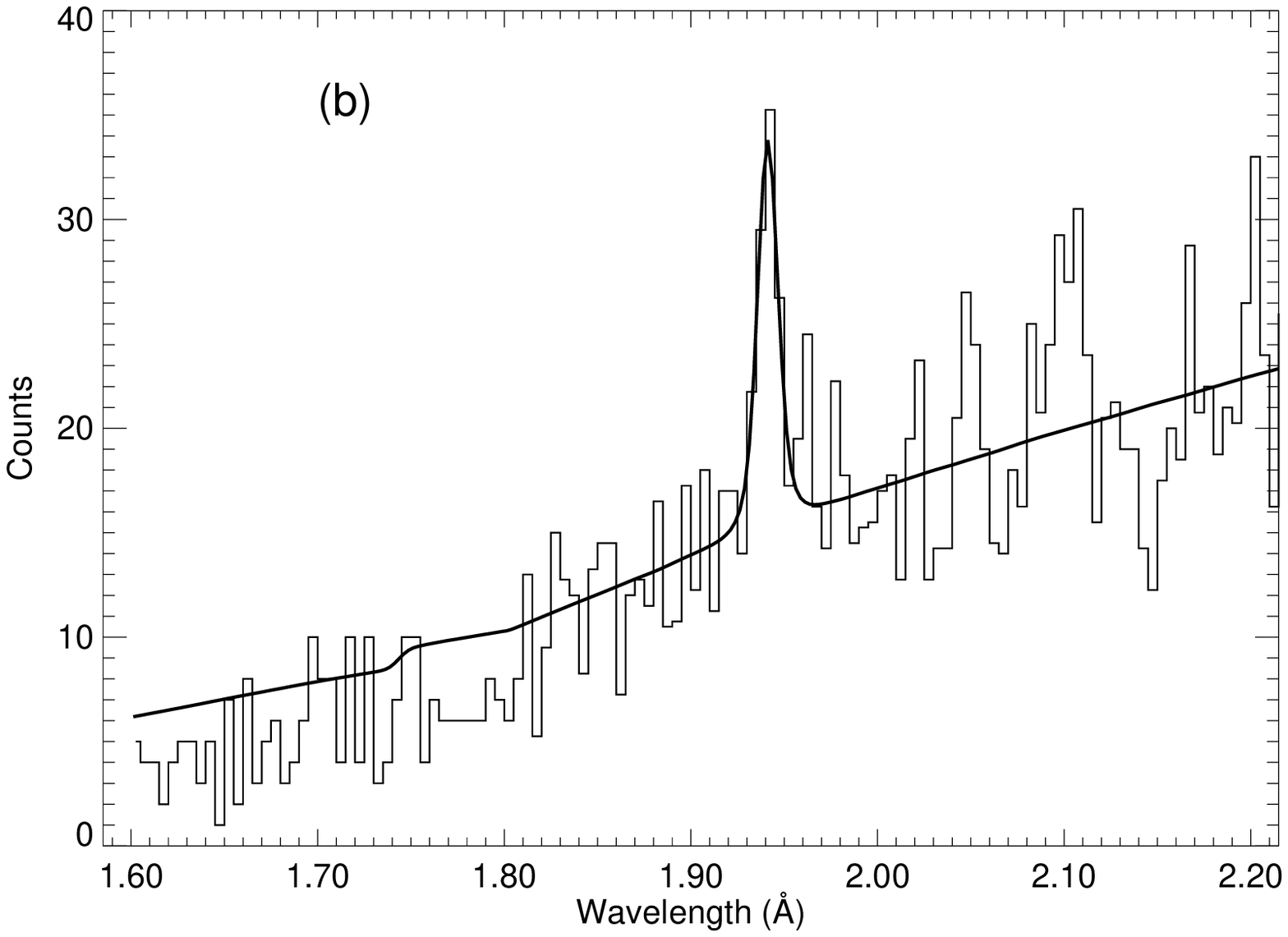}
\end{figure}

\setcounter{figure}{2}
\begin{figure}
\plotone{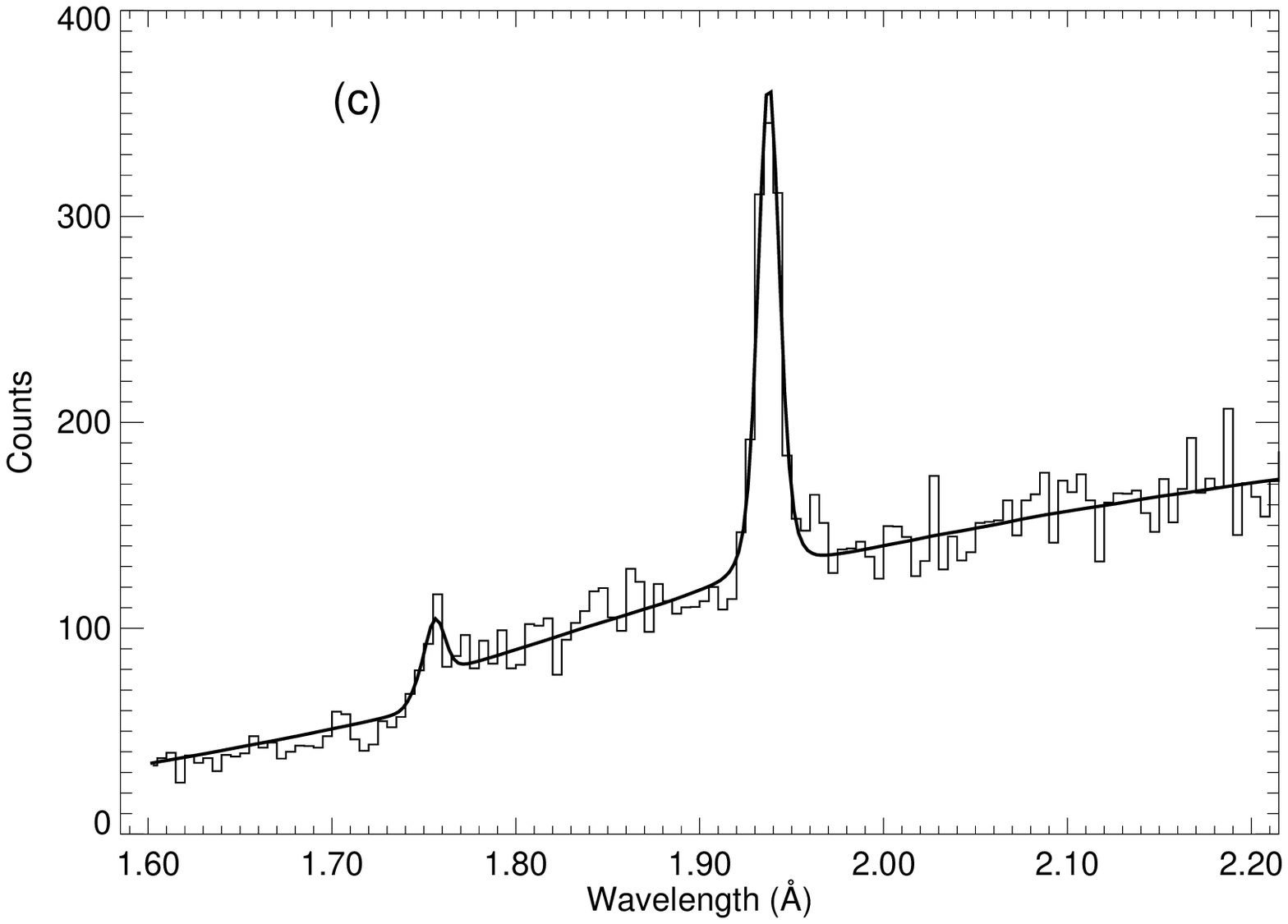}
\end{figure}

\clearpage

\setcounter{figure}{3}
\begin{figure}
\caption{\label{fig:lowenergy}
}
\plotone{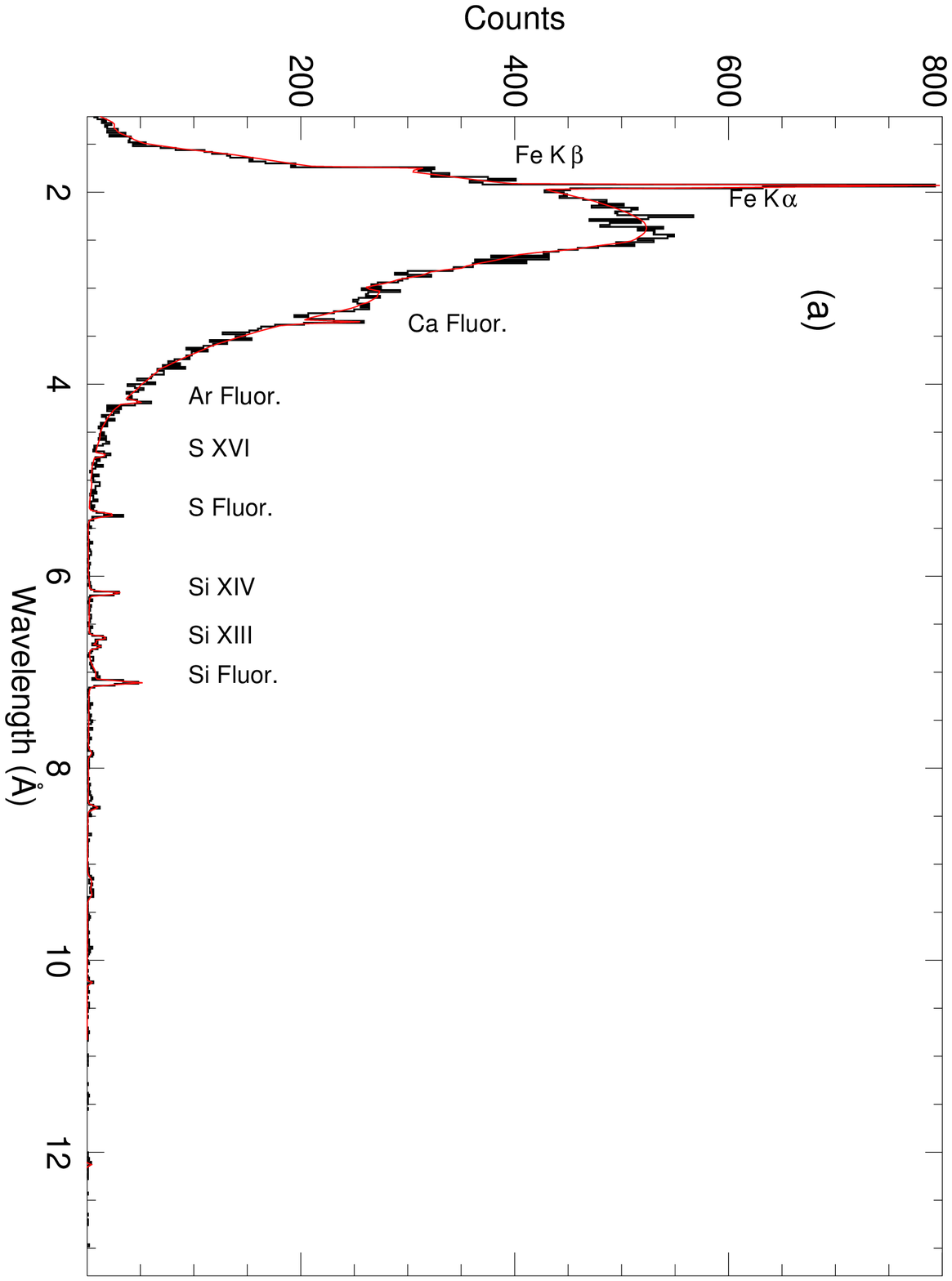}
\end{figure}

\setcounter{figure}{3}
\begin{figure}
\vspace{0.5in}
\plotone{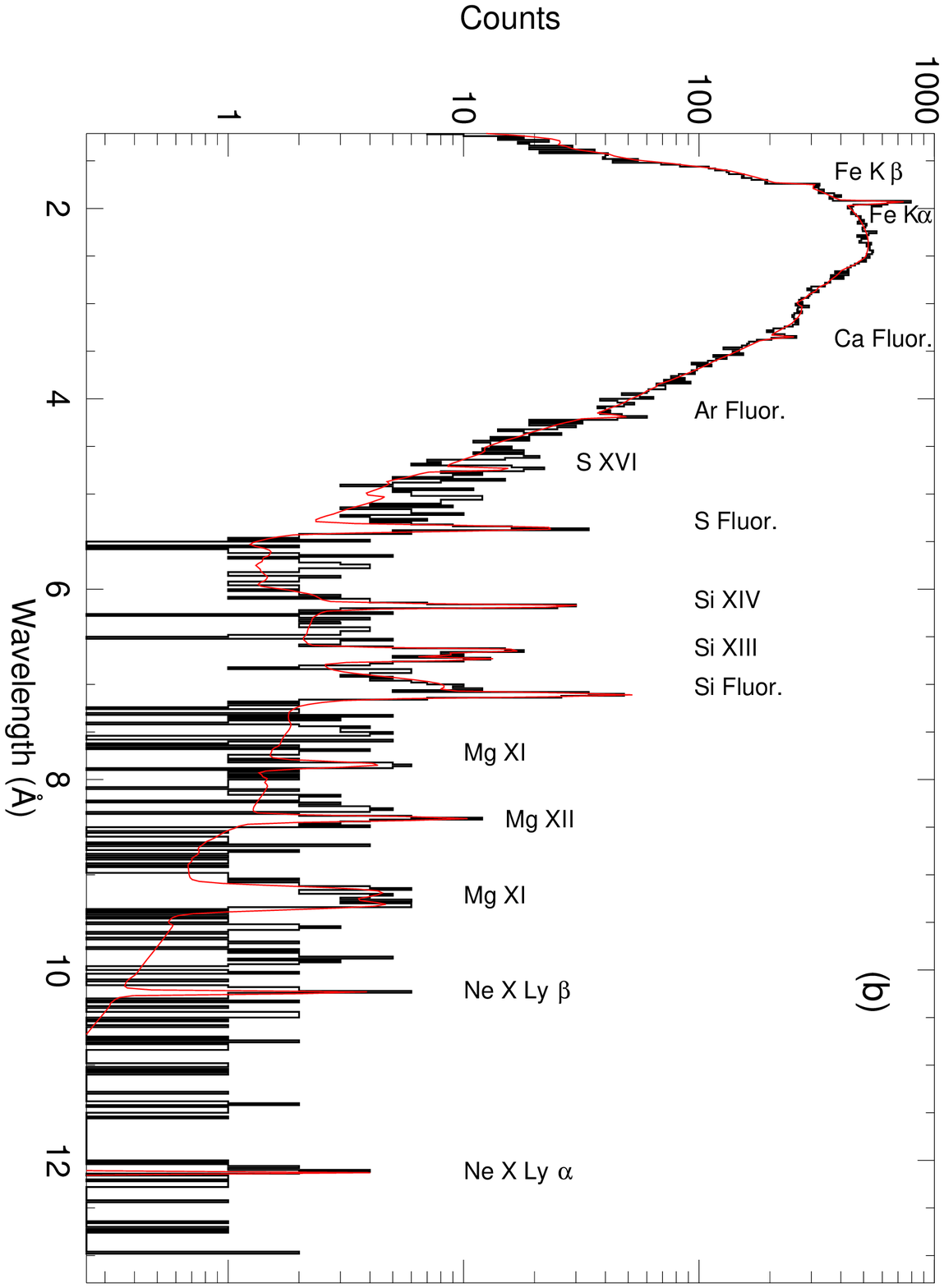}
\end{figure}

\setcounter{figure}{3}
\begin{figure}
\vspace{1in}
\plotone{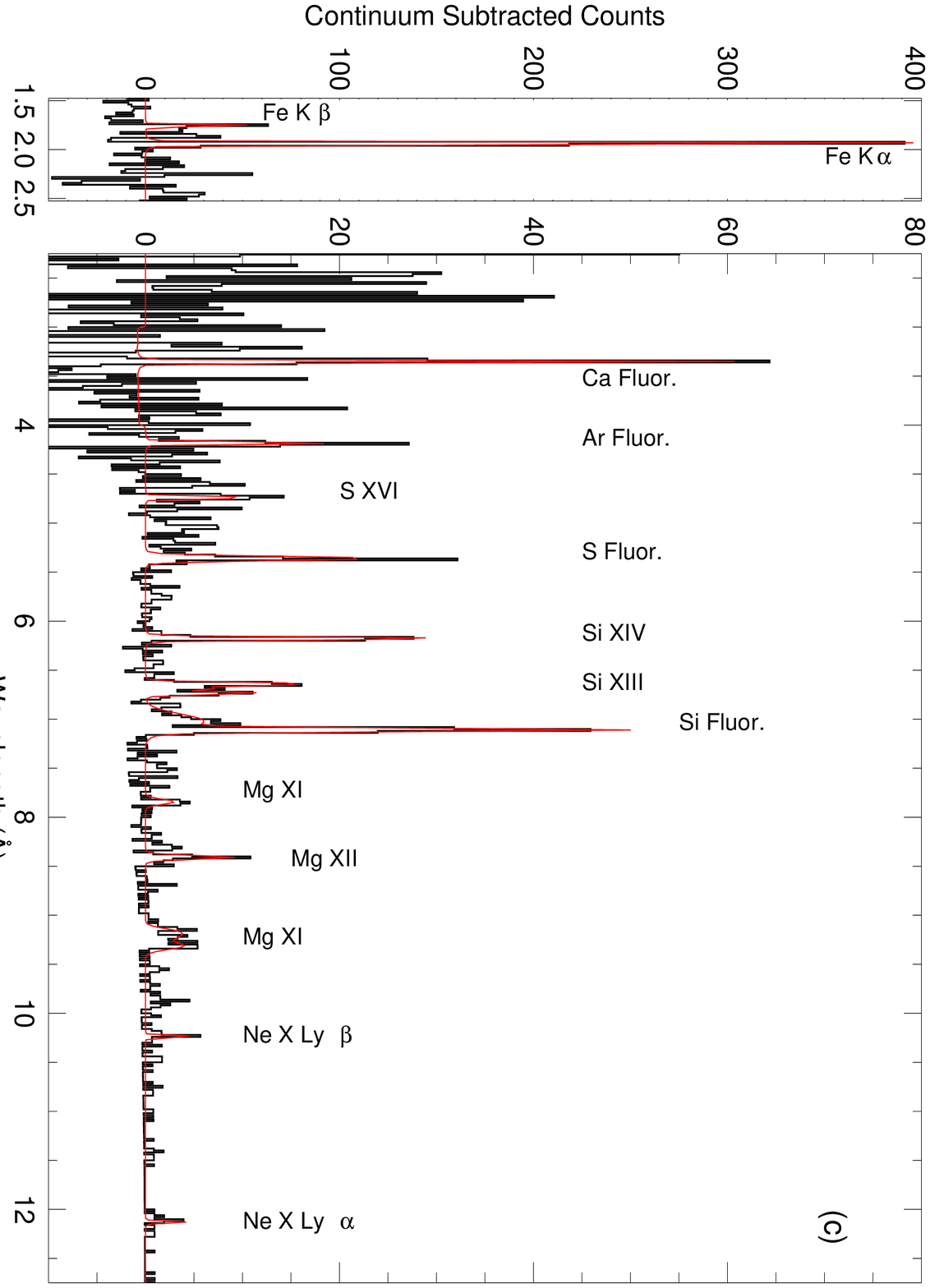}
\end{figure}

\setcounter{figure}{4}
\begin{figure}
\caption{\label{fig:3spectra}}
\plotone{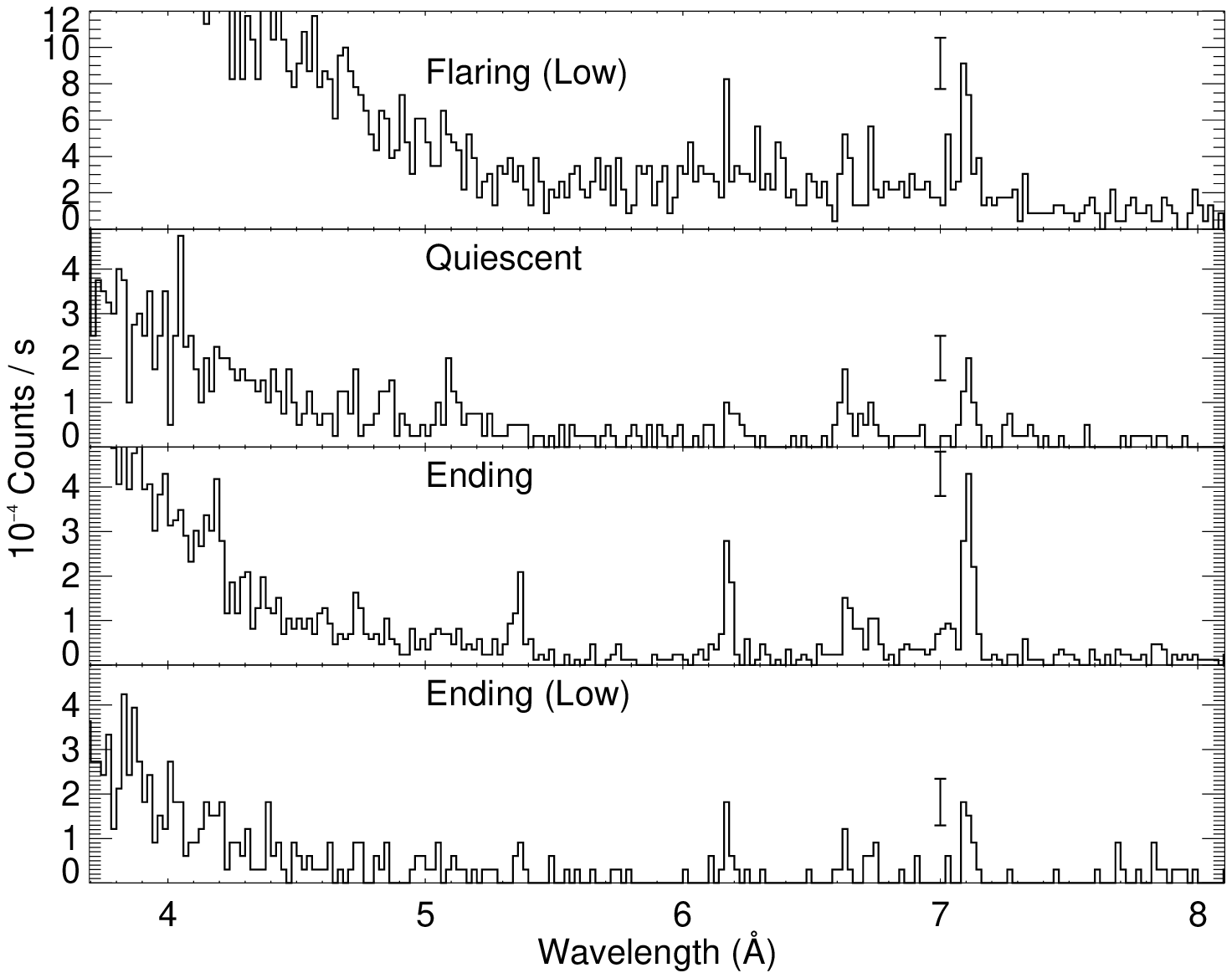}
\end{figure}

\begin{figure}
\caption{\label{fig:corel1}}
\plotone{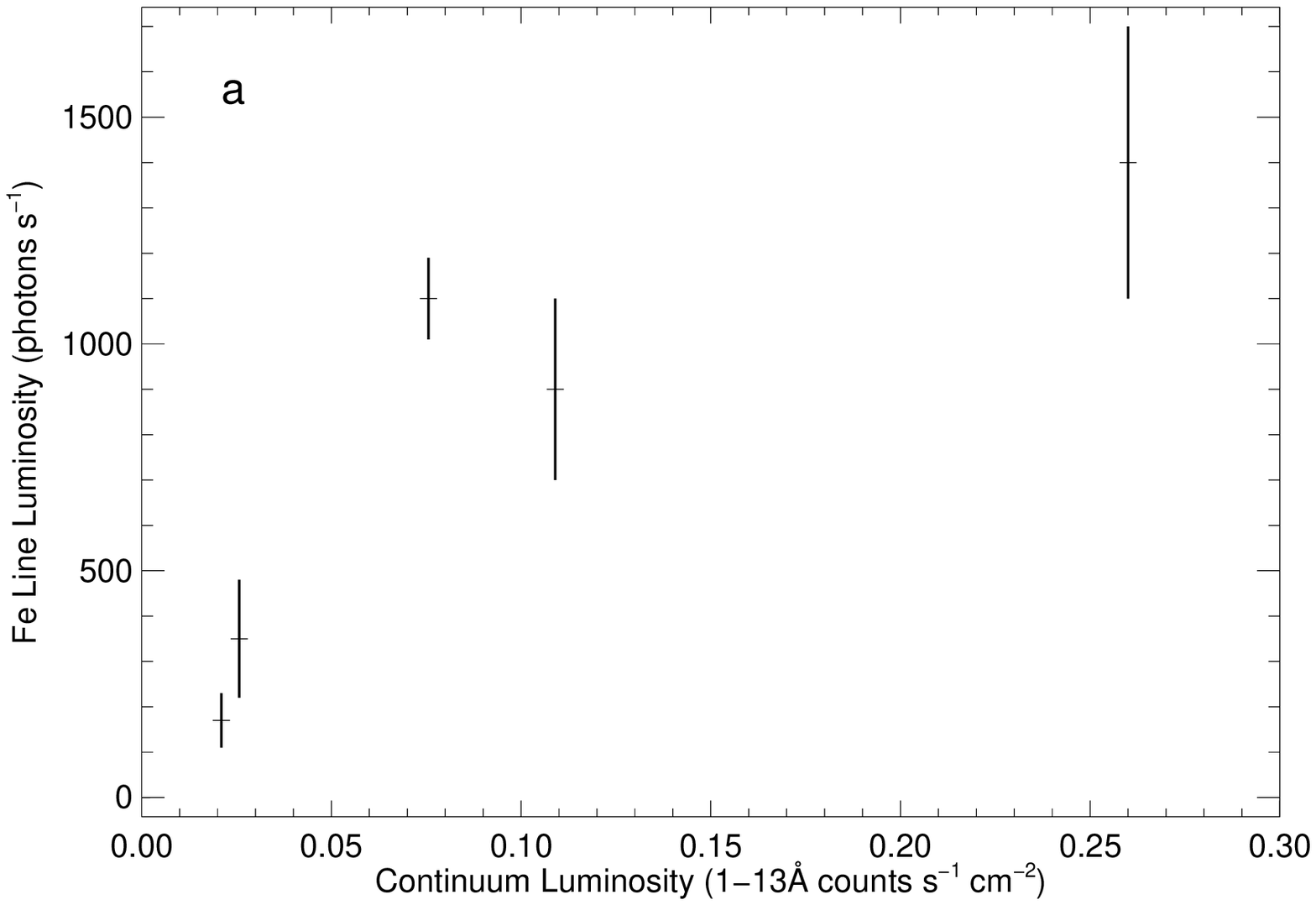}
\plotone{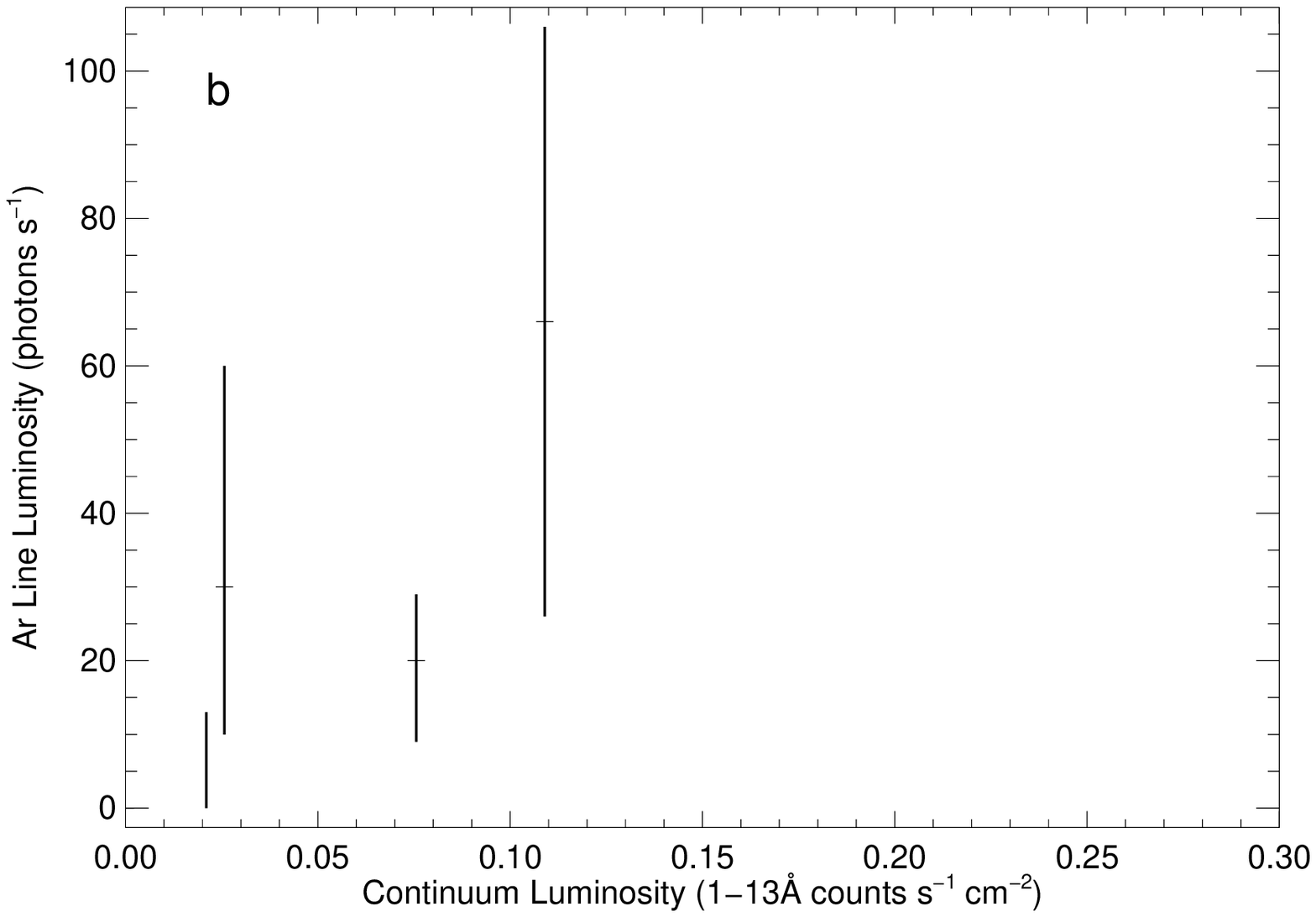}
\end{figure}

\setcounter{figure}{5}
\begin{figure}
\plotone{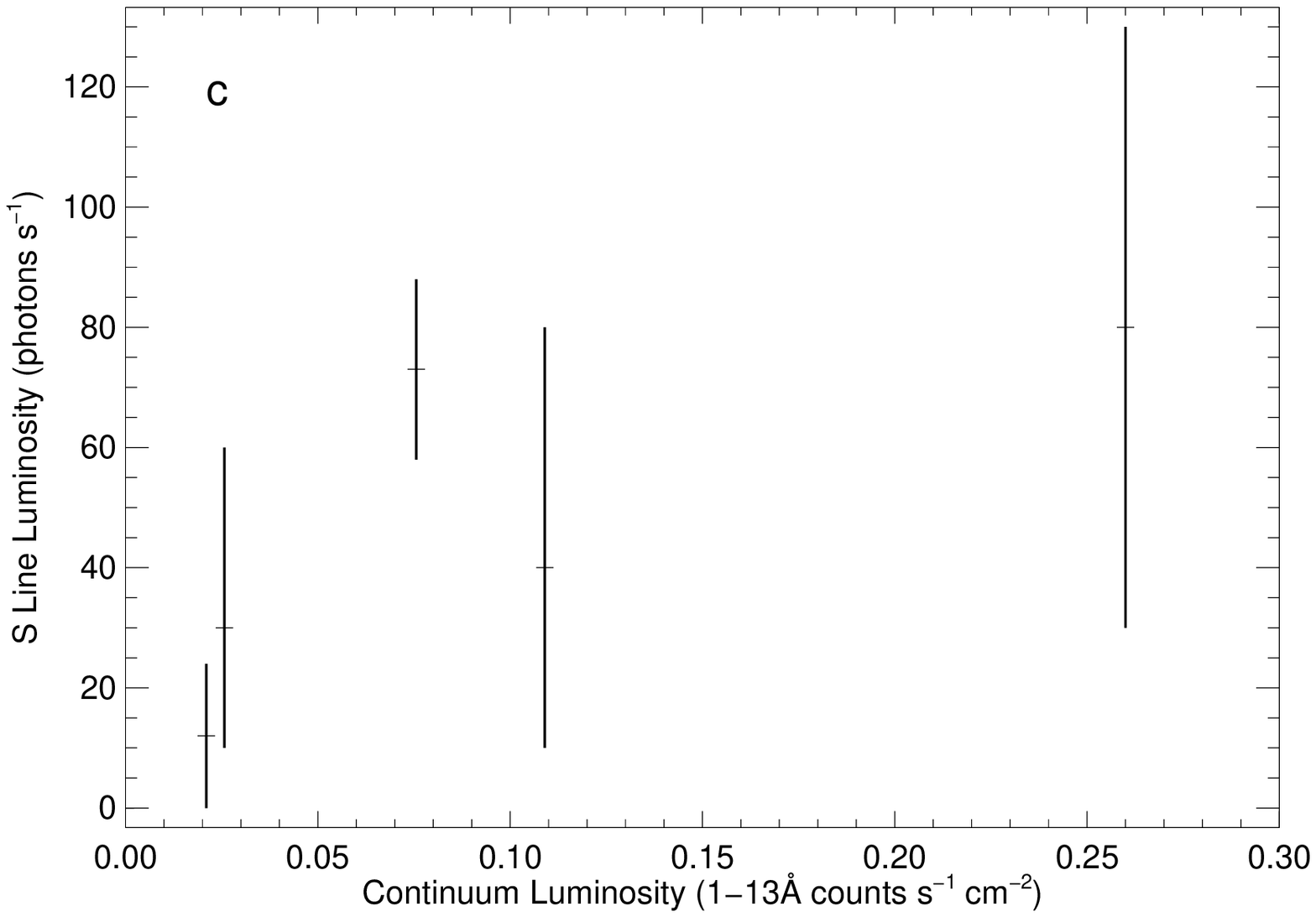}
\plotone{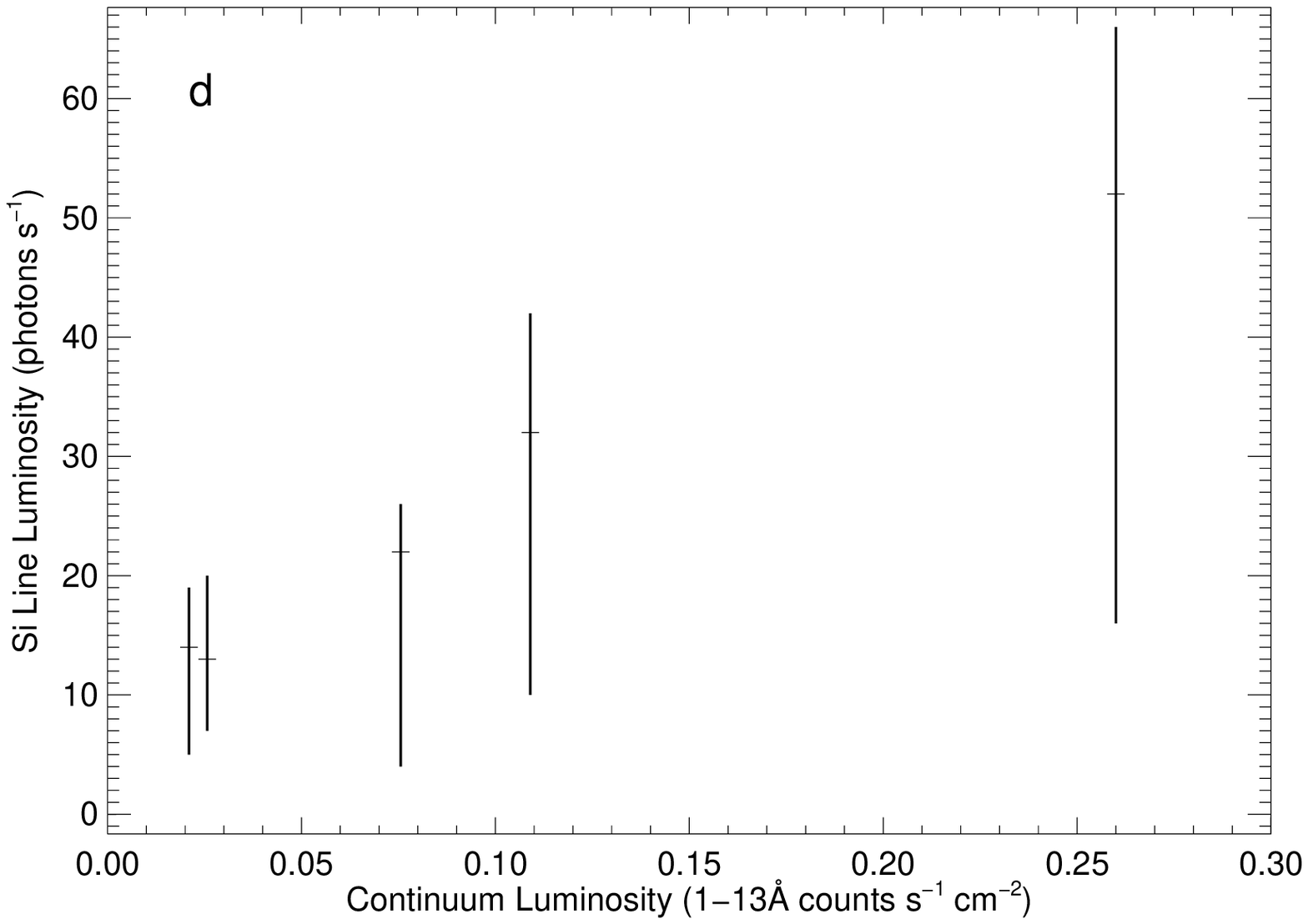}
\end{figure}

\setcounter{figure}{5}
\begin{figure}
\plotone{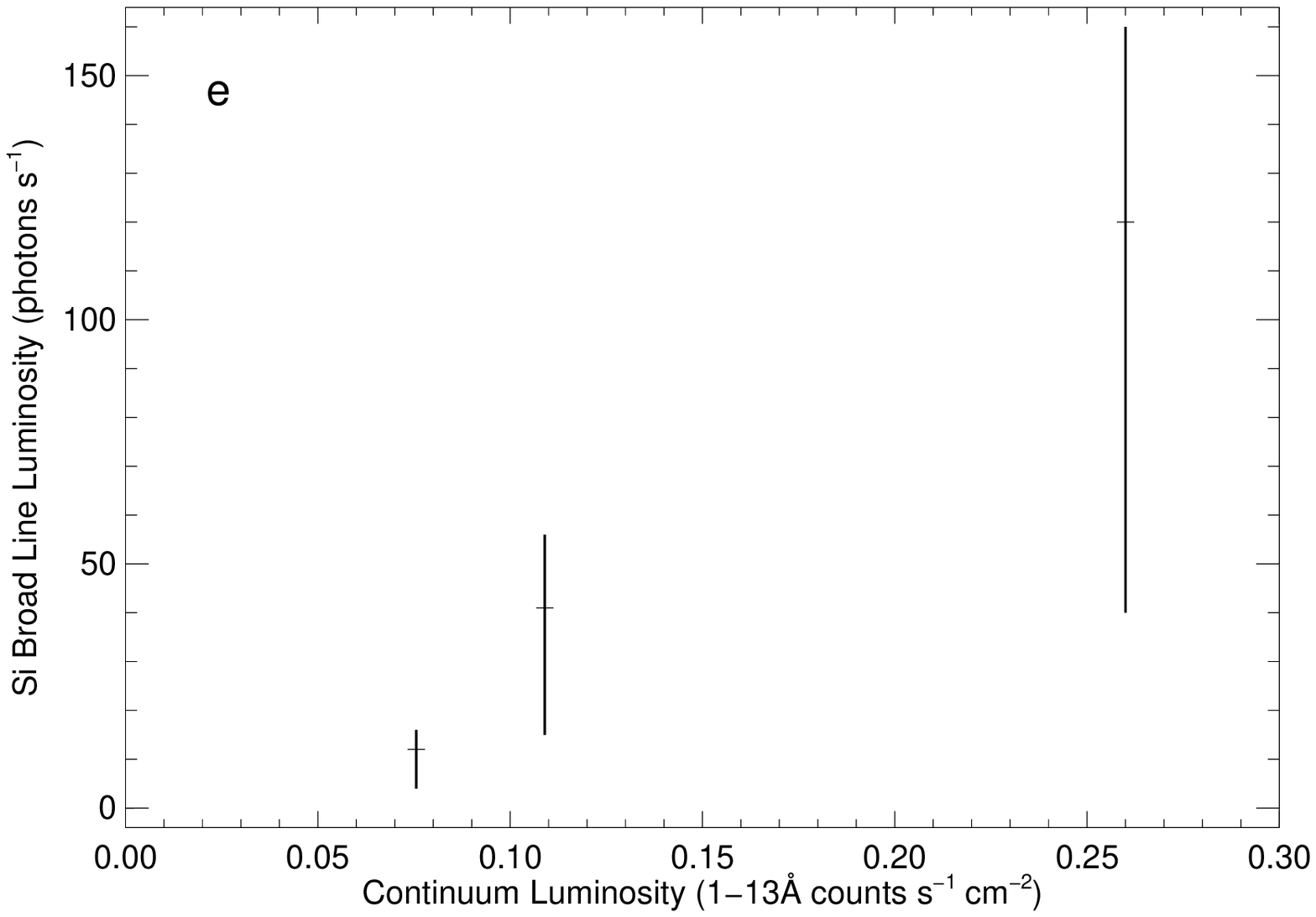}
\end{figure}

\setcounter{figure}{6}
\begin{figure}
\caption{\label{fig:corel2}}
\plotone{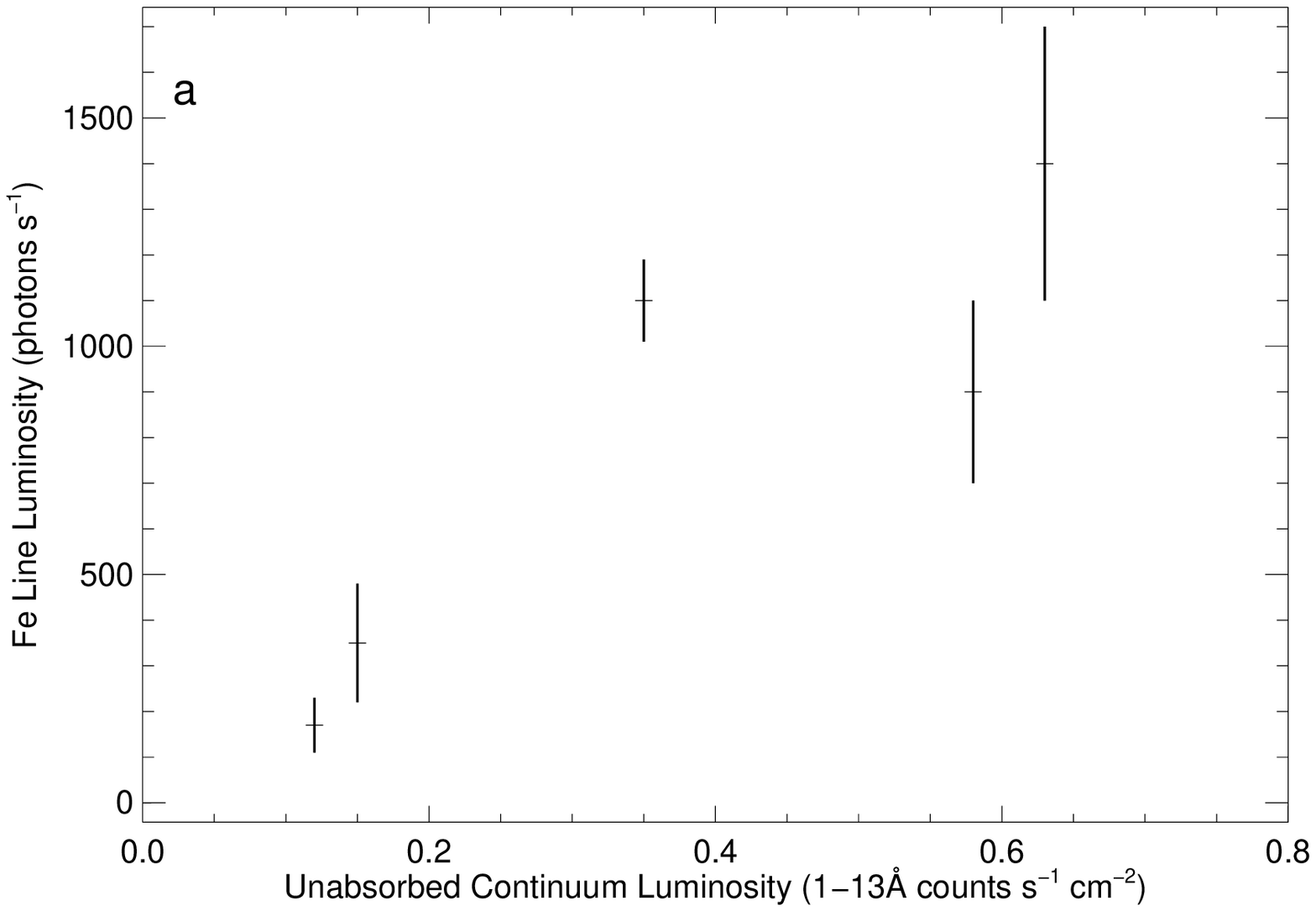}
\plotone{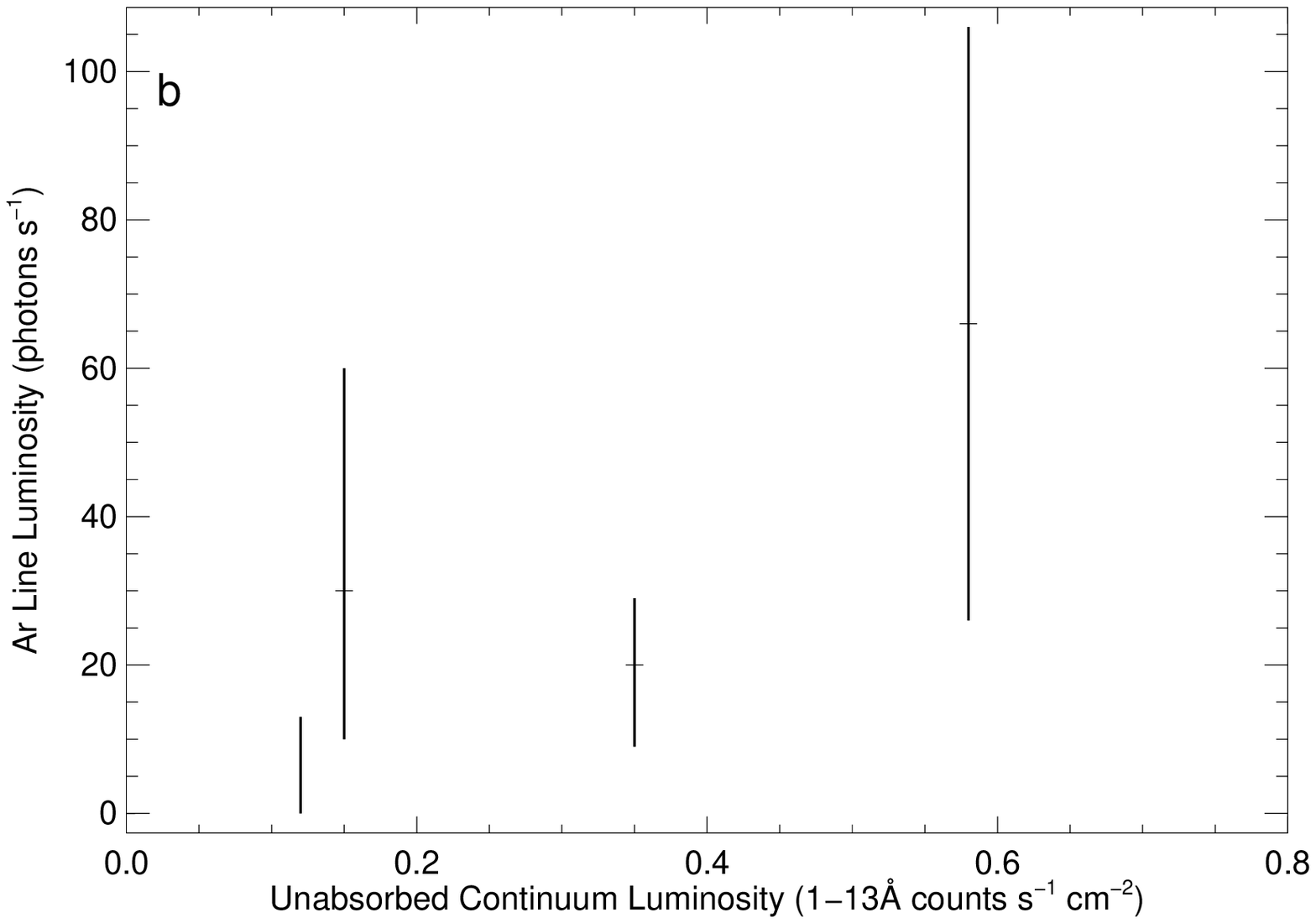}
\end{figure}

\setcounter{figure}{6}
\begin{figure}
\plotone{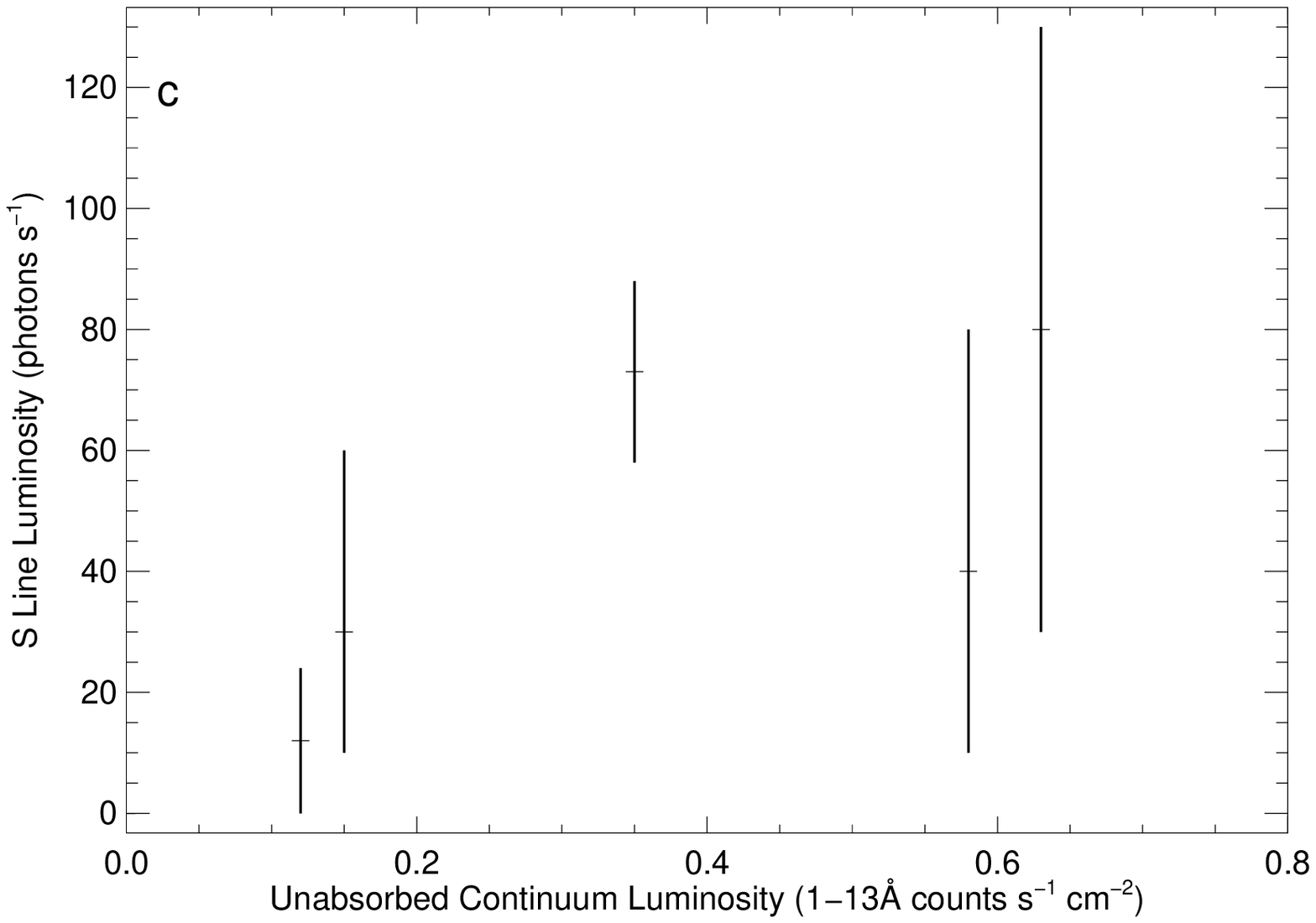}
\plotone{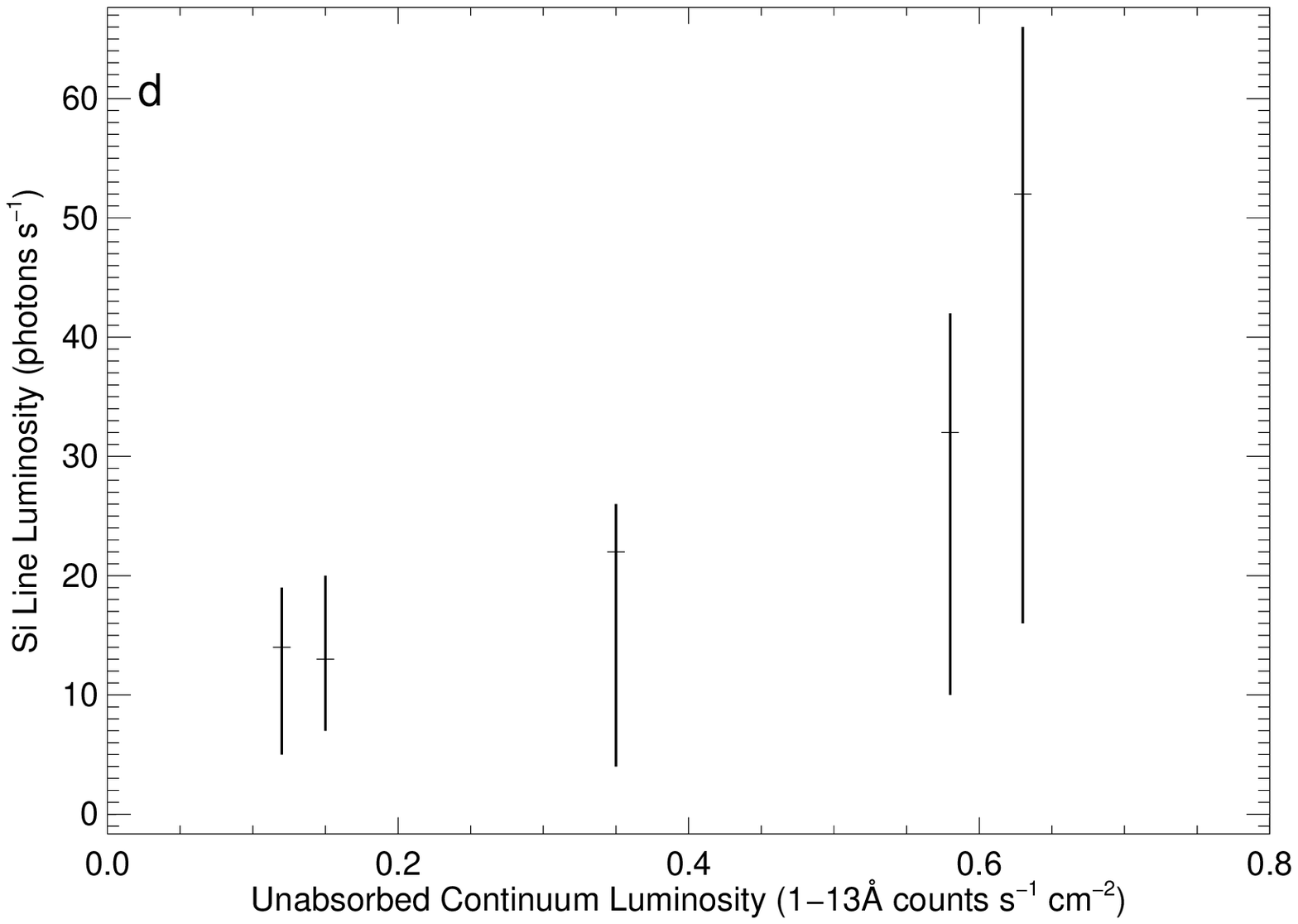}
\end{figure}

\setcounter{figure}{6}
\begin{figure}
\plotone{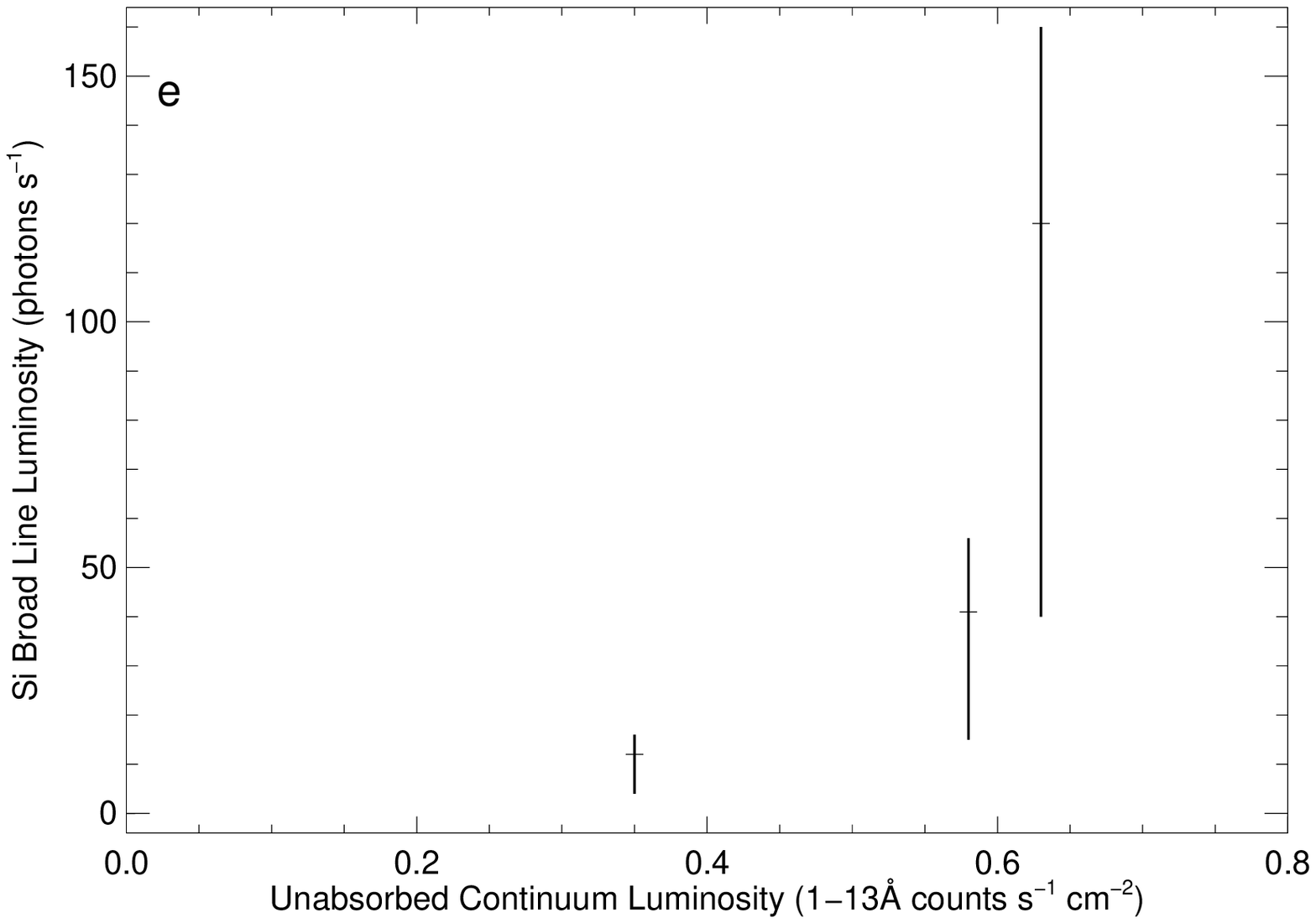}
\end{figure}

\setcounter{figure}{7}
\begin{figure}
\caption{\label{fig:helikesi}}
\plotone{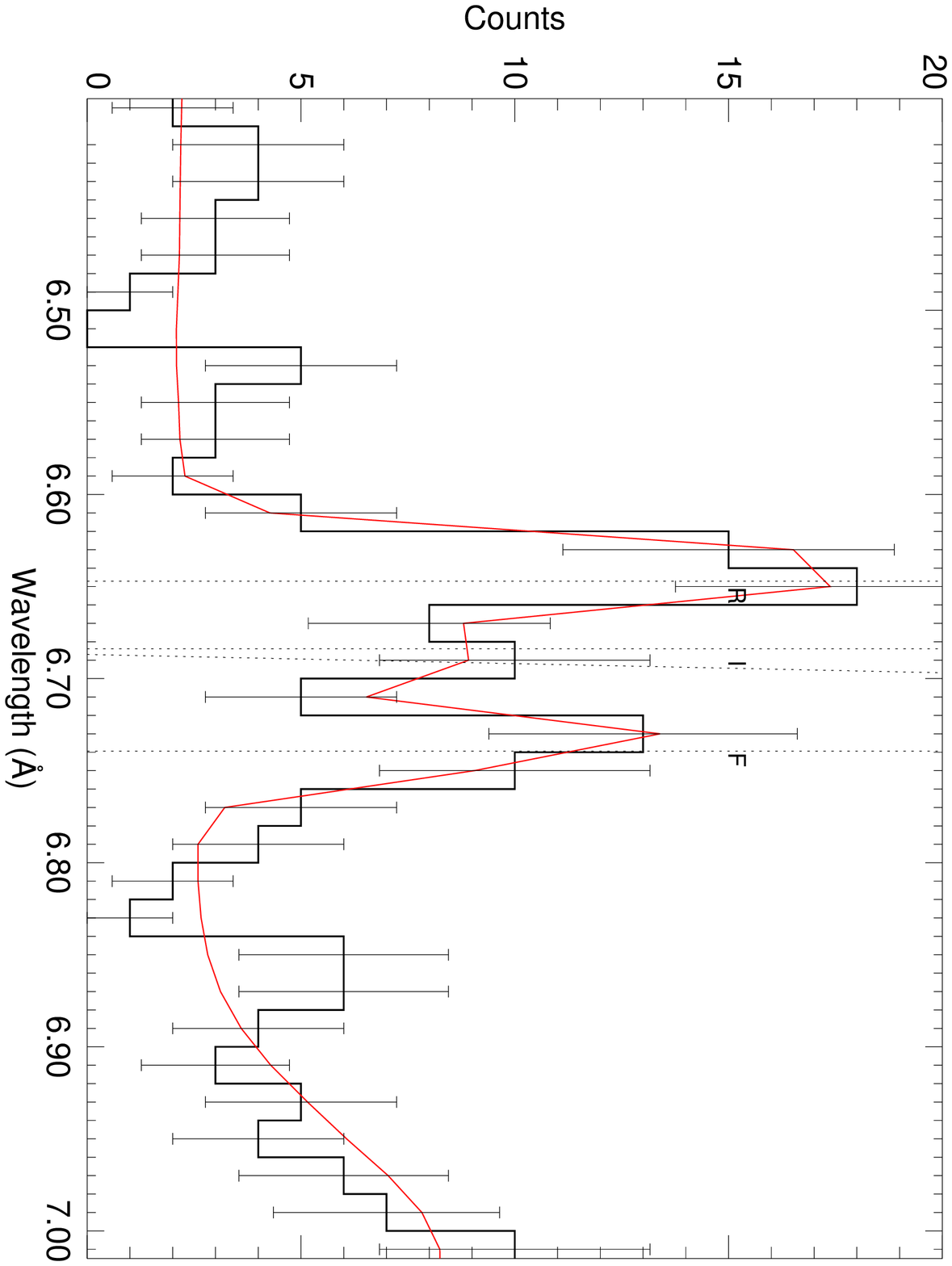}
\end{figure}

\begin{figure}
\caption{\label{fig:allpower}
}
\plotone{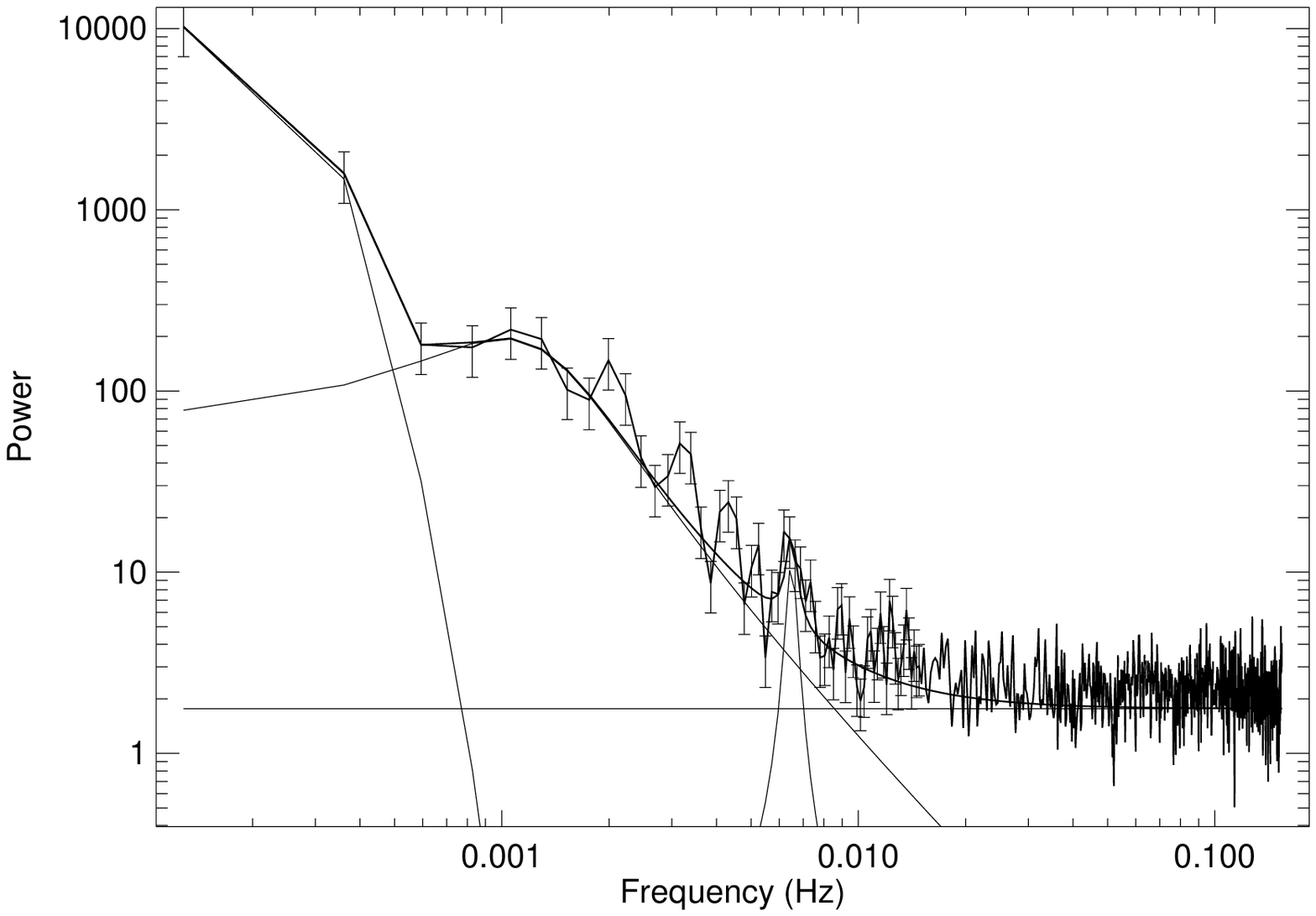}
\end{figure}

\begin{figure}
\caption{\label{fig:linepower}
}
\plotone{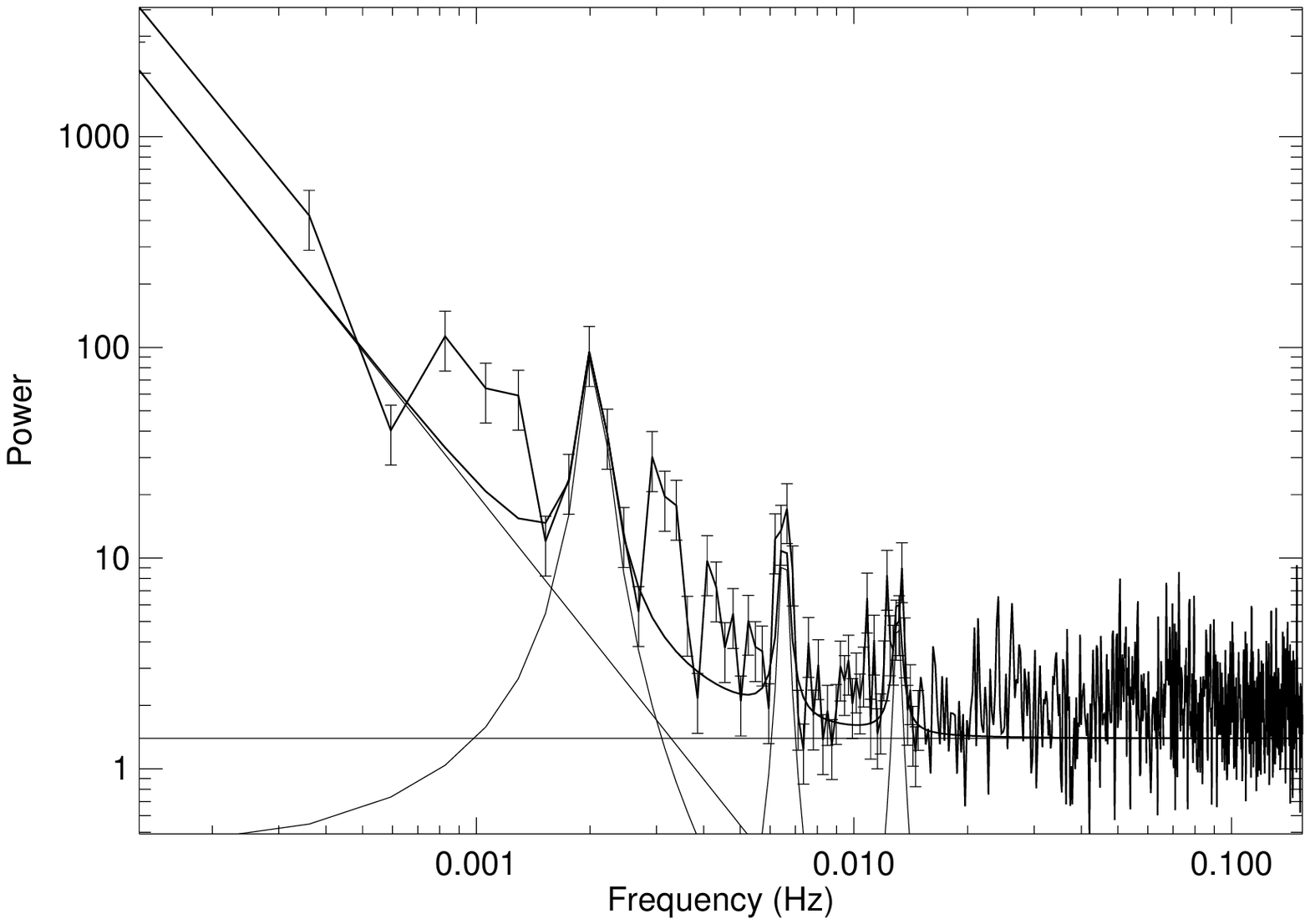}
\end{figure}

\begin{figure}
\caption{
\label{fig:fullpsd}}
\plotone{f11.eps}
\end{figure}

\begin{figure}
\caption{\label{fig:xi}}
\plotone{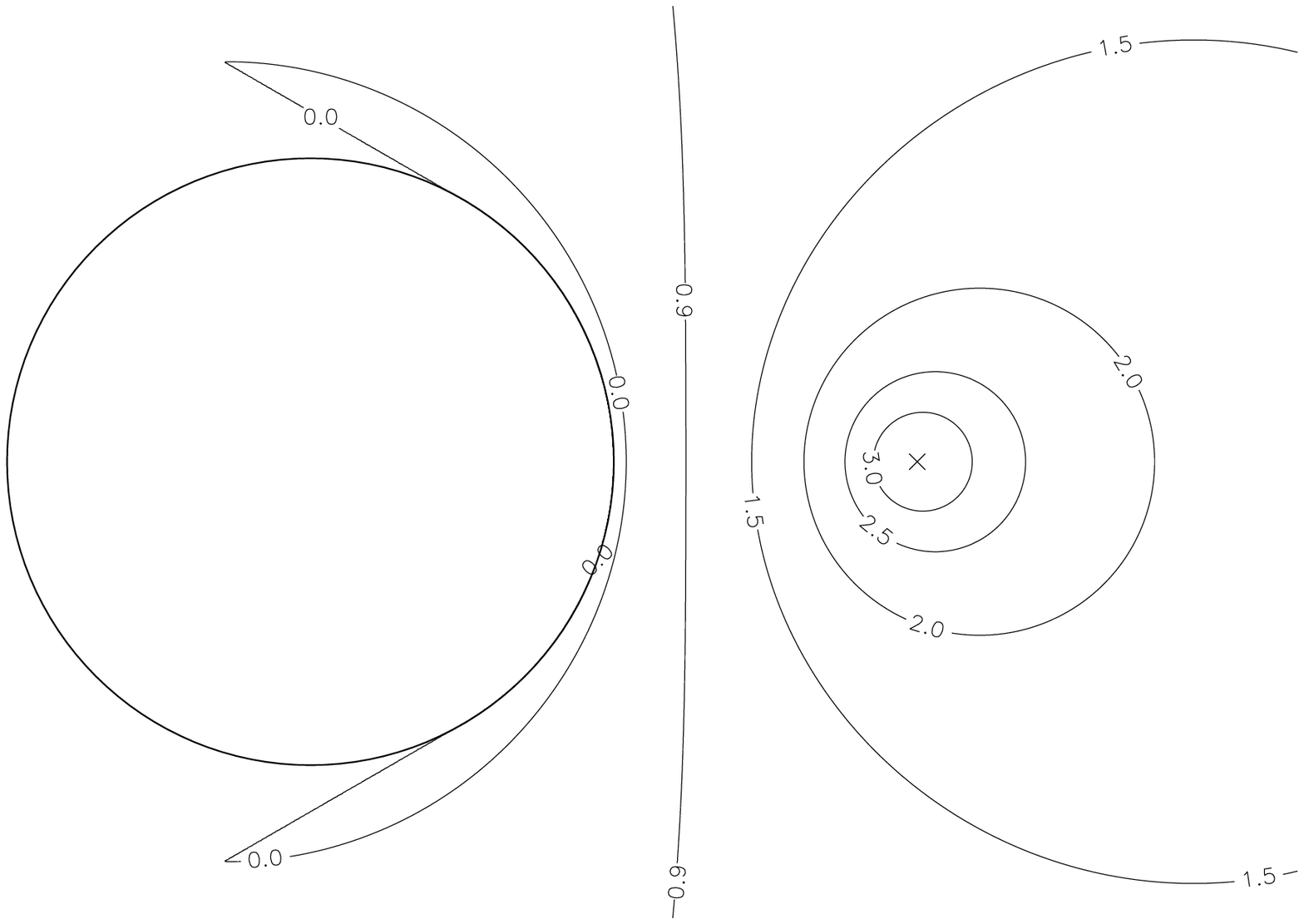}
\end{figure}


\begin{references}

\reference{}Ankay, A., Kaper, L, de Bruijne, J.H.J., Dewi, J., Hoogerwerf, 
R., Savonije, G.J.  2001, A\&A, 370, 170


\reference{} Bautista, M.A., \&\ Kallman, T.R.  2001, ApJS, 134, 139

\reference{}Belloni, T., \&\ Hasinger, G.  1990, A\&A, 230, 103

\reference{} Bhattacharya, D., \&\ Srinivasan, G.  1995, in X-ray
Binaries, ed. Walter H.G. Lewin, Jan van Paradijs, and Edward P.J. van den
Heuvel, Cambridge University Press

\reference{} Blondin, J.M.  1994, ApJ, 435, 756

\reference{} Blondin, J.M., Stevens, I.R., \&\ Kallman, T.R.  1991, ApJ, 
371, 684

\reference{} Blondin, J.M., Kallman, T.R., Fryxell, B.A., \&\ Taam, R.E.
1990, ApJ, 356, 591


\reference{} Bohlin, R.C.  1975, ApJ, 200, 402

\reference{} Bondi, H., \&\ Hoyle, F.  1944, MNRAS, 104, 273

\reference{} Boroson, B., O'Brien, K., Horne, K., Kallman, T., Still, M., 
Boyd, P.T., Quaintrell, H., Vrtilek, S.D.   2000, ApJ, 545, 399


\reference{}Branduardi, G., Mason, K.O., Sanford, P.W.  1978, MNRAS, 185, 
137

\reference{} Brinkmann, W.  1981, A\&A, 94, 323

\reference{} Buff, J., \&\ McCray, R.  1974, ApJ, 188, 37

\reference{} O'Brien, Kieran; Horne, Keith; Kallman, Timothy; Still,
Martin; Boyd, 
Patricia T.; Quaintrell, Hannah; Vrtilek, Saeqa Dil


\reference{} Canizares, C.R., et al.  2000, ApJ, 539, L41

\reference{} Cash, W.  1979, ApJ, 228, 939

\reference{} Clark, J.S., Goodwin, S.P., Crowther, P.A., Kaper, L,
Fairbairn, M., Langer, N., \&\ Brocksopp, C.  2002, A\&A, 392, 909

\reference{} Corbet, R.  1986, MNRAS, 220, 1047

\reference{} Davidson, K., \&\ Ostriker, J.P.  1973, ApJ, 179, 585

\reference{} Doll, H., \&\ Brinkmann, W.  1987, A\&A, 173, 86

\reference{} Fransson, C. \&\ Fabian, A.A.  1980, A\&A, 87, 102 

\reference{} Friend, D.B., \&\ Castor, J.I.  1982, ApJ, 261, 293

\reference{} Gehrels, N.  1986, ApJ, 303, 336

\reference{} Haberl, F., Aoki, T., \&\ Mavromatakis, F.  1994, A\&A, 288, 
796

\reference{} Haberl, F., \&\ Day, C.S.  1992, A\&A, 263, 241

\reference{} Haberl, F., White, N.E., \&\ Kallman, T.R.  1989, ApJ, 343,
409 (HWK)

\reference{} Hammerschlag-Hensberge, G., Howarth, I.D., \&\ Kallman, T.R.  
1990, ApJ, 352, 698

\reference{} Heap, S.R., \&\ Corcoran, M.F.  1992, ApJ, 387, 340

\reference{} Houck, J.~C~\&Denicola, L.~A.\ 2000, ASP Conf.~Ser.~216:
Astronomical Data Analysis Software and Systems IX, 9, 591

\reference{} House, L.L.  1969, ApJSS, 155, 21

\reference{} Huenemoerder, D.P., Canizares, C.R., \&\ Schulz, N.S., 2001, ApJ, 559, 1135

\reference{} Jahoda, K., Swank, J.H., Giles, A.B., Stark, M.J., 
Strohmayer, T., Zhang, W., Morgan, E.H.  1996, SPIE, 2808, 59

\reference{} Kaastra J.S. \&\ Mewe, R.  A\&AS, 97, 443

\reference{} Kallman, T., \&\ Bautista, M.  2001, ApJS, 133, 221

\reference{} Kaper, L, Hammerschlag-Hensberge, G., \&\ Zuiderwijk, E.J.
1994, A\&A, 289, 846

\reference{} Illarionov, A., Kallman, T.R., McCray, R.A., \&\ Ross, R.R.  
1978, ApJ, 228, 279

\reference{} Jimenez-Garate, M.A., Hailey, C.J., den Herder, J.W., Zane, S., 
\&\ Ramsay, G.  2002, ApJ, in press

\reference{} Jones, C., Forman, W., Tananbaum, H., Schreier, E., Gursky, 
H., Kellogg,   E., \&\ Giacconi, R.  1973, ApJ, 181, L43

\reference{} Kallman, T., \&\ White, N.E.  1989, ApJ, 341, 955

\reference{} Langer, S., Ross, R.R., \&\ McCray, R.A.  1978, ApJ, 222, 959

\reference{} Leahy, D.A., Darbro, W., Elsner, R.F., et al.  1983, ApJ, 
266, 160

\reference{} Lee, J., Reynolds, C.S., Remillard, R., Schulz, N.S., Blackman, E.G., 
Fabian, A.C.  2002, ApJ, 567, 1102

\reference{} Levine, A.M., Rappaport, S.A., Zojcheski, G.  2000, ApJ, 541,
194

\reference{} Livio, M., Shara, M.M., \&\ Shaviv, G.  1979, ApJ, 233, 704

\reference{} Marshall, H.L., Canizares, C.R, Schulz, N.S.  2002, ApJ, 564, 
941

\reference{} Miller, J.M., Wojdowski, P., Schulz, N.S., Marshall, H.L.,
Fabian, A.C., Remillard, R.A., Wijnands, R., \&\ Lewin, W.H.G.  2002,
astro-ph/0208463

\reference{} Murakami, T., Kawai, N., Makishima, K., \&\ Mitani, K.  1984,
PASJ, 36, 691

\reference{} Moon, D.-S., \&\ Eikenberry, S.S.  2001a, ApJ, 549, 225

\reference{} Moon, D.-S., \&\ Eikenberry, S.S.  2001b, ApJ, 552, 135

\reference{} Morrison, R., \&\ McCammon, D.  1983, ApJ, 270, 119

\reference{} Nagase, F., Zylstra, G., Sonobe, Takashi, Kotani, T., \&\
Inoue, H.  1994, ApJ, 436, L1

\reference{} Paerels, F., Cottam, J., Sako, M., Liedahl, D., Brinkman, A.C>, 
van der Meer, R.L.J., Kaastra, J.S., Predehl, P.  2000, ApJ, 533, 135

\reference{} Petterson, J.A.  1978, ApJ, 224, 625

\reference{} Porquet, D., Mewe, R., Kaastra, J.S., Dubau, J., \&\ Raassen, 
A.J.J.  2002, Proc. Symposium ``New Visions of the X-ray Universe in the 
XMM-Newton and Chandra Era'', 26-30 November 2001, ESTEC, The Netherlands, 
in press.

\reference{} Pozdnyakov, L.A., Sobol, I.M., \&\ Sunyaev, R.A.  1979, A\&A, 
75, 214


\reference{} Reynolds, A.P., Owens, A., Kaper, L, Parmar, A.N., \&\ Sagreto, 
A.  1999, A\&A 349, 873

\reference{} Ross, R.R., Weaver, R., \&\ McCray, R.A.  1978, ApJ, 219, 292

\reference{} Rubin, B.C., Finger, M.H., Harmon, B.A., Paciesas, W.S., 
Fishman, G.J., Wilson, R.B., Wilson, C.A., Brock, M.N., Briggs, M.S., 
Pendleton, G.N., Cominsky, L.R., Roberts, M.S.  1996, ApJ, 459, 259

\reference{} Sako, M., Liedahl, D.A., Kahn, S.M., \&\ Paerels, F.  1999,
ApJ, 525, 921

\reference{} Saraswat, P., Yoshida, A., Mihara, T., Kawai, N., Takeshima,
T., Nagase, F., Makishima, K., Tashiro, M., Leahy, D.A., Pravdo, S., Day,
C.S.R., Angelini, L.  1996, ApJ, 463, 726

\reference{} Schulz, N.S., Cui, W., Canizares, C.R., Marshall, H.L., Lee,
J.C., Miller, J.M., \&\ Lewin, W.H.G., 2002a, ApJ, 565, 1141

\reference{} Schulz, N.S., Canizares, C.R., Lee, J.C., \&\ Sako, M.
2002b, 
ApJ, 564, L21

\reference{} van Loon, J. Th., Kaper, L., \&\ Hammerschlag-Hensberge, G.  
2001, A\&A 375, 498

\reference{} Weisskopf, M.C., O'Dell, S.L., \&\ van Speybroeck, L.P.  
1996, 
Proc. SPIE, 2905, 2

\reference{} Weisskopf, M.C., Brinkman, B., Canizares, C., Garmire, G.,
Murray, S., \&\ Van Speybroeck, L.P.  2002, PASP, 114, 1

\reference{} Wojdowski, P.S., Liedahl, D.A., Sako, M., Kahn, S.M., \&\ 
Paerels, F.  2002, ApJ, submitted

\end{references}
 \end{document}